\RequirePackage[hyphens]{url}
\PassOptionsToPackage{square,numbers}{natbib}
\documentclass[%
 reprint,
nofootinbib,
 amsmath,amssymb,
 aps,
]{revtex4-1}

\usepackage[T1]{fontenc} 
\usepackage[utf8]{inputenc}
\usepackage{listings}
\usepackage[most]{tcolorbox}

\usepackage{hyperref}  

\usepackage{orcidlink}
\usepackage{longtable}
\usepackage{enumitem}
\usepackage[export]{adjustbox} 
\usepackage{graphicx} 
\usepackage{csvsimple} 
\usepackage{float} 
\restylefloat{table} 
\usepackage{booktabs} 
\usepackage{subfiles} 

\hypersetup{
    colorlinks=true,
    linkcolor=blue,
    filecolor=magenta,      
    urlcolor=cyan,
    pdftitle={Overleaf Example},
    pdfpagemode=FullScreen,
    }

\usepackage{amsthm}
\usepackage{thmtools}

\usepackage [english]{babel}
\usepackage [autostyle, english = american]{csquotes}
\MakeOuterQuote{"}

\makeatletter
\newcommand{\reformatNOTE}{\def\thmt@space##1(##2){##2}}
\makeatletter

\declaretheoremstyle[
  headfont=\sffamily\bfseries,
  bodyfont=\normalfont,
  headformat={\NAME\ \NUMBER: \NOTE},
]{ex}

\declaretheorem[
  preheadhook=\reformatNOTE,
  style=ex,
  parent=subsection,
  name={Network Attribute},
]{attribute}

\declaretheorem[
  preheadhook=\reformatNOTE,
  style=ex,
  sibling=attribute,
  name={Paper-mill Attribute},
]{papermillattribute}

\begin{document}

\title{Identifying Fabricated Networks within Authorship-for-Sale Enterprises}

\author{Simon J. Porter \orcidlink{0000-0002-6151-8423}0000-0002-6151-8423 }
\author{Leslie D. McIntosh \orcidlink{0000-0002-3507-7468}0000-0002-3507-7468}

\affiliation{Digital Science,London, GB}

\date{January 2024}

\begin{abstract}
Fabricated papers do not just need text, images, and data, they also require a fabricated or partially fabricated network of authors.  Most `authors' on a fabricated paper have not been associated with the research, but rather are added through a transaction. This lack of deeper connection means that there is a low likelihood that co-authors on fabricated papers will ever appear together on the same paper more than once. This paper constructs a model that encodes some of the key characteristics of this activity in an `authorship-for-sale' network with the aim to create a robust method to detect this type of activity.  A characteristic network fingerprint arises from this model that provides a robust statistical approach to the detection of paper-mill networks. The model suggested in this paper detects networks that have a statistically significant overlap with other approaches that principally rely on textual analysis for the detection of fraudulent papers. Researchers connected to networks identified using the methodology outlined in this paper are shown to be connected with 37\% of papers identified through the tortured-phrase and clay-feet methods deployed in the Problematic Paper Screener website. Finally, methods to limit the expansion and propagation of these networks is discussed both in technological and social terms.
\end{abstract}

\maketitle

\section{Introduction}
\label{sec:1}

The study of paper mills---\textit{the organised manufacture of falsified manuscripts that are submitted to a journal for a fee on behalf of researchers} \cite{p6}, has become a research topic that is quickly establishing its importance in the safeguarding of the integrity of the research process.  Indeed, with ongoing technological developments, research in this area is set to play a critical role in safeguarding the integrity of the scholarly record, and may even prove to be existentially important. 

It is estimated that 2\% of all journal submissions across all disciplines originate from paper mills \cite{p6}, both creating significant risk that the body of research that we rely on to progress becomes corrupted, and placing undue burden on the submission process to reject these articles. By understanding how the business of paper mills---the technological approaches that they adopt, as well as the social structures that they require to operate---the research community can be empowered to develop strategies that make it harder, (or ideally) impossible for them to operate.

Most of the contemporary work in paper-mill detection has focused on identifying the signals that have been left behind inside the text or structure of fabricated papers that result from the technological approaches that paper mills employ. Current efforts to automate the detection of suspicious research include: the identification of `tortured phrases' in text that are designed to mask plagiarism \cite{10.48550/arxiv.2210.04895}; the identification of common phrases and templates that appear to be used by specific paper mills \cite{10.1038/d41586-022-04367-z}; the detection of manipulated images \cite{10.1128/mbio.00809-16}; the manipulation of data \cite{10.26508/lsa.202101203,10.1002/1873-3468.13747}; and, the creation of falsified data through statistical methods \cite{10.1136/bmj.331.7511.267, 10.31234/osf.io/bxau9}.  These efforts to detect fabricated papers are in constant tension with the technical capabilities of paper-mill operators, with the latest generative AI techniques being explored on both sides\cite{10.1038/d41586-023-01780-w, 10.31219/osf.io/wk3g7}. 

To date there has been little research into whether the social structures that enable paper mills to flourish also leave an imprint on the fabricated papers that they create. Research into the social structures that foster scientific misconduct, of which paper mills are only a part, has focused on the systemic drivers within research culture \cite{10.1371/journal.pone.0127556}. It is argued \cite{10.1089/ees.2016.0223} that undesirable behaviours associated with paper mills are driven by the precise systems put in place to ensure that research is well-regulated. In recent years these systems have led to highly competitive, metricized and unhealthy approaches rewarding production frequency and volume. Other pressures include requirements to publish in order to be promoted within a clinical career, or meet the requirements of a doctoral program \cite{p6}.  Whilst these insights shine a torch on the cultural changes that need to occur to reduce the need for paper mills, they do not in themselves lead to actionable insights that can be used to detect paper-mill output. 

We assert that a closer inspection of the social drivers behind paper-mill output \textit{can} be translated into insights that can be used directly to dismantle paper-mill enterprises.  Specifically, we believe that practices fundamental to the business models of paper mills leave a collaboration fingerprint that can be identified using techniques familiar to the `Science of Science' \cite{10.1126/science.aao0185}. These fingerprints are difficult to mask without significant social engineering or outlandish payments to bona fide academics, both of which may price paper mills out of the market, and hence should remain a high-quality mechanism for detection even as technological fabrication techniques evolve.

Our key insight is that paper mills do not just fabricate content, convincing paper-mill outputs also fabricate co-authorship networks of the researchers that they involve. Of the estimated 2\% of all journal submissions across all disciplines made up by paper-mill submissions, most are usually submitted by authors who have not previously submitted to the journal or who have not previously published in an academic venue \cite{p6}. Papers with these two attributes already carry an increased risk profile compared with an average paper. Armed with this information, a journal review process could adapt their processes to apply a higher level of scrutiny to these submissions. Paper-mill papers that successfully make it through the submission and review processes are likely to require a more convincing fabricated author network. 

By understanding and identifying the properties of these fabricated co-authorship networks, we propose effective strategies aimed at identifying paper-mill activity. By identifying authors that are very likely to be part of fabricated co-authorship networks, we provide techniques that can make it easier to detect paper-mill output during the submission process; highlight journals that appear to be compromised; and reduce the supply of willing participants by identifying 'at risk' researchers within universities. 

This paper is arranged as follows: In the remainder of this introductory section we formulate a model of co-authorship activity that allows us to develop a set of conjectures for how these activities could manifest in a co-authorship network. We then discuss strategies for validating the appearance of these characteristics through comparison with established, known paper-mill networks. In Section~\ref{sec:2} we describe both the technical and mathematical approach that we have used to implement the model and explore the characteristic described in the current section. In Section~\ref{sec:3} we present an analysis of these methods and compare them with a set of validation steps proposed in the prior section.  Finally, in Section~\ref{sec:4} we make recommendations on how to apply the results of this approach both at the institutional and publisher level.

\subsection{Principal attributes of networking exhibiting paper-mill activity}
\label{theory:authorship-for-sale}
Research collaborations typically arise from social situations such as shared research group membership, PhD mentoring relationships, or conference and seminar attendance, but this is not how paper-mill co-authors form relationships and the differences in the genesis of these different network leaves a fingerprint. In this section we review the fingerprints left by paper-mill activities in the co-authorship network and encapsulate these as network attributes that we can use to identify the specific strategies used by paper mills.

In the case of an ordinary, non-paper-mill-generated manuscript, the research network behind the paper has emerged organically: when we read such a multi-authored paper, we do not just encounter new ideas and experiments, we read the work as the product of a collaboration that has taken time to build---often representing researchers from many different career stages. New research arises out of existing research communities, connected through time, via a network of supervisions and collaborations. New PhD students are trained by supervisors, who likely become co-authors before branching off into new collaborations forged through projects and conferences. 

Normal, organically-generated research networks typically build, connect, and spread slowly early in a career, accelerating in later years. It is not unreasonable to image that there exist characteristic levels of connectedness as research careers develop, constrained perhaps by a research network equivalent of a Dunbar number (the maximal number of friends that a person can reasonably call friends given the time and cognitive load of maintaining friendships) \cite{10.1016/0047-2484(92)90081-j}. Co-author network shapes may differ by field of research or location, among other factors, but they have more similarities than differences. Outliers may exist, for example a younger researcher from riding the coat-tails of a prolific senior researcher, but through co-author affiliations can deduce certain types of relationship and hence explain certain categories of prolific or highly-connected work. Occasionally, true outliers exist---tremendously brilliant or lucky individuals who have happened to work on a new stream of research in atypical ways, and this is why the processes that we elucidate here should not be 100\% automated - our methods are designed to highlight outliers, but they do not give the reason behind their outlier status.

Networks involving ``ordinary'' fraudulent research activity such as plagiarism, data/image fabrication and ghost authorship also develop organically `inside the tent', or endogenously, within established research networks. These endogenous networks also tend to be localised and build over time, with the machinery and knowledge of misconduct retained within the team. While the impact of internal misconduct of this nature is undeniably damaging to our ability to trust science, they are limited as the penalties for such actions often end careers \cite{y9j,eg}. 

Such behaviour is naturally localised as, if it becomes known outside a fairly tight network of individuals, then the likelihood of a report to an institutional ethics board, funder oversight committee, editor, publisher or professional organisation becomes too high and the consequences noted above are much more likely to ensue. While these types of infractions are damaging to the scholarly record, they are perhaps more corrosive to trust both among academics and between the public and research as a publicly-funded enterprise. More directly and practically, funding is potentially diverted to the wrong places (both people and fields), inappropriately disadvantaging ethical researchers who may be passed over in favour of a more prolific but dishonest colleague. 

However, paper-mill networks are different to both the other types that we have discussed above. We can regard this type of network as being `outside-the-tent', or exogenous to existing research networks.  In these cases, individual researchers either willingly depart, or are enticed or entrapped into departing, from their usual local networks and purchase authorship-for-sale positions on publications from paper mills. In doing so, they become part of a global research misconduct network. 

This network does not have the same constraints as either natural, locally occurring networks or the misconduct networks described above--it is unconstrained by the number of publications it can produce, or by the number of researchers it can recruit. In addition to its structural ability to scale, we propose that global misconduct networks associated with authorship-for-sale paper-mill outputs can be distinguished from organic research networks based on the following attributes: 

\begin{attribute}[The majority of researchers within authorship-for-sale networks will tend to have a `young' publication age]
\label{att:young_researchers}
We define the quantity of ``publication age'' to be the elapsed time from an author's first publication to their most recent publication. Most authors on paper-mill papers---at least the authors that have paid to be there---will tend to be early in their career (as measured by publication age). Researchers with established careers have reputations to protect. 

The motivations identified by the Committee on Publication Ethics (COPE) for purchasing authorship on a paper are also associated with young researchers. These include \cite{p6}:
\begin{enumerate}
    \item Doctoral students being unable to graduate unless they have published a paper
    \item A clinician in a hospital required to publish before they can apply for or be eligible for a promotion
   \item A researcher trying to boost their publication profile in order to appear more accomplished so that they can secure a research grant
\end{enumerate}

The exception to this are `foundation authors' (covered shortly in Attribute~\ref{att:foundation_authors}), who are involved in authorship-for-sale publications for other reasons. 
\end{attribute}

\begin{attribute}[High-volume researchers within the authorship-for-sale network should have an egocentric network with a low clustering coefficient]
\label{att:low_clustering_coefficient}
In contrast to the organic research network, the network of researchers associated with paper-mill papers will not be a product of an evolving collaboration. Most ‘authors’ on a fabricated paper will have been associated transactionally (i.e., through the chance purchasing of an author position on the same paper), with a low likelihood that they will ever appear together on the same paper again. Furthermore, if a person purchases an author position on a paper, and does this many times (perhaps in building a profile in order to obtain a grant), the network they form will not reflect an evolving research collaboration of a young researcher. Instead, the created network will look like a highly centralised research hub. Remove the hub, and the network will fall apart into the sub-clusters created by individual publications. This pattern of research collaboration for researchers with a ‘young’ publication age may not be uncommon to a few types of researchers (e.g., statisticians), however, it is unusual for most disciplines.

More formally, for a researcher who is heavily engaged with authorship-for-sale networks, their personal network (their egocentric network) of co-authors will have a low clustering coefficient \cite{10.1038/30918}, with the only connection between the different papers being the researcher themselves. The clustering coefficient is defined to be the residual density of the researcher's egocentric network after the removal of their (central) node. This measure is a good proxy for determining the centrality of a researcher to their local community.  One caveat here is that this method works poorly in cases where publications include a large number of researchers such as high-energy particle physics.
\end{attribute}

\begin{attribute}[An authorship-for-sale co-authorship network will have a limited number of senior `foundation' authors]
\label{att:foundation_authors}
As a group of authors on a paper with little publication history is likely to invite extra scrutiny, an effort must be made to fabricate or co-opt researchers with more substantial publication records. It is reasonable to assume that most researchers with established research networks will not have anything to do with fabricated papers. The high degree of competition for tenured research positions, and the public nature of research outputs ensures that academic careers cannot be based on high volume, low-quality paper-mill output. To be discovered is to damage your career \cite{ptw}.

Agents employ multiple means to fabricate co-authors with publication histories:

\begin{enumerate}
    \item Authors can be added to papers without their knowledge (this carries a high risk of discovery) \cite{vk};
    \item Given that authors' correspondence is handled by the paper mill for authorship-for-sale transactions \cite{p6}, we speculate that authors from previously accepted paper-mill papers could be reused (without permission or recourse for them to complain);  
    \item Authors with peripheral academic careers (with less to lose) could be co-opted into the process;
    \item Authors could be co-opted for financial advantage in order to offset the risk of discovery \cite{10.1002/leap.1574}.
\end{enumerate}

We can regard each of these attributes as resulting from a strategy (with an accompanying business model) of the paper mill in service of profitable false papers production. Each attribute or strategy can be used alone or in combination with others.  However, as we have described, each strategy leaves a fingerprint in the academic co-author network. 

With the exception of the (high-risk) strategy associated with the first attribute, each of these strategies takes effort to develop. Either profiles must be created based on previously submitted work, or relationships (coercive or otherwise) must be developed. As ‘foundation authors’ provide credibility to a paper, their presence will be required on most fabricated papers. This tension in effort in developing foundation authors and the need to include them on as many paper-mill papers as possible creates the likelihood that these assets will be overused. The result of overused foundation authors is that these researchers will end up with an artificially inflated per-year number of publications over a short space of time. The egocentric research networks that these authors create will also have a low clustering coefficient (Attribute~\ref{att:low_clustering_coefficient}).

Of the attributes described above, with their associated production strategies, we conjecture that an approach focused on building foundation authors from previously submitted authorship-for-sale papers is the most scalable, as it offers opportunities to continue create profiles, even as the pool of willing participants from the organic research network becomes (potentially) exhausted. As these profiles are constructed from paper-mill output, they are also likely to be associated with researchers that also have a reasonably young publication age.

\end{attribute}

\begin{attribute}[High-volume participants in the authorship-for-sale cohort should form a network of low cluster coefficient egocentric networks]
\label{att:authorship-for-sale-netowrk}

Counterintuitively, although each high-volume participant in an authorship-for-sale network will have a low clustering coefficient for their egocentric network, all high-volume participants should be loosely connected to a number of other high-volume participants via random co-authorship ‘collisions’ on authorship-for-sale papers. The sum of these connections should form a connected graph that is largely separate from the organic research network.

A further property of authorship-for-sale graphs is that their development takes place over a significantly shortened timescale than would be natural for an organically developed graph. For paper-mill authors, it isn't time that gradually connects researchers together, but rather the exogenous factor of the scale of the authorship-for-sale provider. 
\end{attribute}

\begin{attribute}[Researchers within the authorship-for-sale network will exhibit low levels of mentorship]
\label{att:mentorship}

Egocentric co-authorship networks that arise mainly from patronage to authorship-for-sale papers will not contain patterns of mentorship such as dominant relationships between PhD researchers and their advisors and the collaborative networks to which they are introduced through their community. Although it is not possible to identify advisors and supervisors from their publishing profiles, researchers can be identified by publication age. Within an authorship-for-sale co-authorship network then, we do not expect to find mentorship pairings. 

\end{attribute}

\begin{attribute}[Papers within the authorship-for-sale network will likely have a greater number of authors than the discipline norm]
\label{att:author_numbers}

As paper-mill activity is a fundamentally commercial activity and hence driven by profit, it follows that efficiency is also a driver.  This leads to a further artifact in the network: The more author ``slots'' that can be sold on a manuscript, the greater profit margin for that manuscript since the paper only needs to be produced and published once. Rebalancing so that the ratio of paid authors to real authors is not in the short-term interest of profitability. This property has been observed empirically in an investigation of a Russian paper mill, where it was found that paper-mill papers exhibited an elevated average of 3.9 authors per paper, compared with an overall discipline average of 2.6 authors per paper \cite{10.1002/leap.1574,10.1016/j.joi.2020.101110}.  

\end{attribute}

\subsection{Secondary attributes of networking exhibiting paper-mill activity}

In the last section, we restricted our attention to how the attributes of co-authorship graphs were affected by paper-mill strategies. In this section, we summarise two further effects that we see in the data, but which do not relate to the co-authorship network. 

\begin{papermillattribute}[Papers within the authorship-for-sale network should form a citation cartel]
\label{att:citation_cartel}
As the nature of paper-mill papers is that they tend to contain either low quality or repetitious work, they are highly unlikely to attract citations from mainstream research. As a result, it is reasonable to expect that citations to paper-mill papers will come from other paper-mill papers.  
\end{papermillattribute}

\begin{papermillattribute}[Evidence of peer-review enablement]
\label{att:peerreview}
Authorship-for-sale networks also require peer review processes that allow the publication of papers with questionable quality. Like foundational authors, complicit peer reviewers are likely to be a relatively scarce resource. As we know, peer reviewers take time to be inserted into a journals trusted network of peer reviewers.  This means that where open peer review data are available, it is possible to track the behaviour of referees and evaluate their actions with respect to known paper-mill papers.
\end{papermillattribute}

\subsection{Brokered authorship-for-sale as a complicating factor}

While we have tried to define robust attributes aligned with data fingerprints that identify the practices of paper mills, paper mills do not always generate their own content. For example, authorship-for-sale also occurs in the context of research papers involve researchers who have carried out their research in good faith.  This is possible due to the research processes information asymmetry \cite{10.1093/jleo/ewp031} that can exist on papers when it comes to authorship. Not all authors on a paper may be aware of the exact contributions of other authors. Perhaps only the lead author might. This asymmetry leaves the door open for an unscrupulous lead author to add non-contributing authors to a paper facilitated by a paper mill without other legitimate authors being aware. In this example, research is carried out and written up in good faith by some of the authors on a paper can still end up being brokered by a paper mill \cite{10.1002/leap.1574}.

In these cases, although the contributing authors' research networks will look ‘normal,’ the networks of the researchers who have purchased authorship spots are still likely to have a low cluster coefficient. 

The presence of brokered authorship-for-sale articles means the authorship-for-sale research network will be connected to the organic research network. This in turn means that not all researchers that are connected to an authorship-for-sale network will have done something wrong. Use of any methodology that identifies suspicious author practice in a way that would be detrimental to an author must also require individual investigation.

\section{Methodology}
\label{sec:2}

\subsection{Datasources}
To analyse network patterns in the literature we used \textit{Dimensions} from Digital Science  \cite{10.3389/frma.2018.00023}. \textit{Dimensions} is well-positioned for the analysis that we perform here as:

\begin{enumerate}[label=(\alph*)]
\item \textit{Dimensions}' inclusion criteria is based on an output having a unique identifier rather than on an editorial policy \cite{10.3389/frma.2018.00023} and hence does not implicitly limit the range of publication venues that can be analysed. This approach means that the full bad-actor network is available for analysis and, in particular, authorship-for-sale networks can be tracked across all publication venues;

\item \textit{Dimensions} is frequently used in paper-mill analyses, notably tortured phrases analyses (see, for example, \cite{10.48550/arxiv.2210.04895}). As at Oct 2023, links to \textit{Dimensions} have been included in 3952 posts on the social network for public research review -  PubPeer\footnote{At the time of publication, PubPeer can be searched for use of Dimensions via the URL https://pubpeer.com/search?q=*app.dimensions*} \cite{10.7551/mitpress/11087.003.0015};

\item All of the \textit{Dimensions} data is represented as a dataset on Google Cloud Platform on BigQuery \cite{10.3389/frma.2021.656233}, facilitating the global inspection of trends, and allowing the easy addition of other external datasets \cite{10.3389/frma.2022.835139}, such as Retraction Watch \cite{q,zqc}, ORCiD \cite{Montenegro.2023}, and tortured phrases \cite{10.48550/arxiv.2210.04895}.
\end{enumerate}

\subsection{Detecting unusual collaboration signals in the scholarly record}
\label{method:unusualsignal}
To attempt to detect a ‘paper-mill co-authorship signal' in the literature, we first establish network shapes associated with authorship-for-sale that can be identified. Based on our theorised shape of an ‘authorship-for-sale’ researcher (see Section~\ref{theory:authorship-for-sale}), these network shapes should exhibit a low clustering coefficient, where most of the connections between authors go through the researcher around which the egocentric network is constructed (Attribute \ref{att:low_clustering_coefficient}). Importantly, in order to be positively identified in our analysis, at least a subset of these networks need to be rare enough to be distinguished from normal patterns of research collaboration.

One tool that we use extensively is the concept of ``publication age'', which we defined in Attribute~\ref{att:young_researchers}.  Through basic calculations, we access this quantity by using the researcher disambiguation data in \textit{Dimensions}.  Each researcher in \textit{Dimensions} is mapped to an identity, which is, wherever possible, linked to an ORCiD.  As we have access to disambiguated profiles of researchers, we assess the date of the first paper that we know to be associated with that profile in \textit{Dimensions} and calculate the separation between that date and the date of the most recent paper associated to the same researcher profile. It should be acknowledged that the researcher profiles in \textit{Dimensions} are not completely accurate as the process for their creation is, in part, statistical in nature and relies on data availability. This means that publication age can only be used statistically in our work.  However, the profiles, data and calculated ages are sufficiently robust as to allow meaningful analysis. 

\subsection{Describing network shapes}
\label{methods:networkshape}
This methodology also relies on the researcher disambiguation described above. In creating the graphs under consideration, we use the fact that each individual researcher in \textit{Dimensions} is identified by a unique researcher ID. We define the shape of a researcher's immediate co-authorship network to be the depersonalised collection of edges and nodes for a given publication year.  We call this the network ``shape'', as it is unconcerned with who the collaborating researchers are, what institutions they are at, where papers have been published or any other identifying material. The "shape" is merely the underlying structure of the collaborative graph that is important. We then parameterise these collections of edges and nodes by the calculation of two quantities: 

\begin{enumerate}[label=(\alph*)]
    \item The number of researchers (nodes) in their network;
    
    \item The clustering coefficient $C$ of the network calculated as the number of edges $E$ in the network divided by the number possible edges as a function of the number of nodes $N$ when the central researcher is removed:
    
    \begin{equation}
        C = \frac{2(E-N)}{N(N-1)},
    \end{equation} 
    
    where N = total number of nodes -1,  and E = total number of edges  -N.
\end{enumerate}
        
For each shape, we can then use the two quantities above to assign a uniqueness measure across the dataset. This means that every author-year combination had assigned to it the two quantities above and the frequency of those quantities was compared across all such author-year pairings. To calculate the research network shape, only journal articles of type research article were used (e.g, no commentaries, reviews). Publications with greater than twenty authors were also excluded to avoid distortions that publications with high numbers of authors create in local co-authorship networks.

Researchers were also allocated approximate career stages based on their publication age in five-year increments (Table \ref{tab:career_stages}). Throughout this analysis career, stage labels are used for convenience, although the actual status of an individual researcher within a cohort may differ in reality.

\begin{table}
\centering

\begin{tabular}{|l|| l  |l |} \hline 

 \textbf{Career Stage}&\textbf{Example}& \textbf{Publication years} \\ \hline 

 I&Student & 0-4 \\ \hline 

 II&Postdoc & 5-9 \\ \hline 

 III&Early career & 10-14 \\ \hline 

 IV&Established & 15-19 \\ \hline 

 V&Career Building & 20-24 \\ \hline 

 VI&Peak Production & 25-29 \\ \hline 

 VII&Advanced & 30-34 \\ \hline 

 VIII&Senior Researcher & 35-100 \\ \hline

\end{tabular}
\caption{Approximate career stages based on publication age. Examples are used indicative, although the actual status of an individual researcher within a cohort may differ in reality.}
\label{tab:career_stages}
\end{table}

The model we constructed through the attributes listed in the previous section suggests that rare network shapes should be associated with researchers with a ‘young’ publication age (Stages I and II) (Attribute~\ref{att:young_researchers}). Further, as our profile of ‘authorship-for-sale’ researchers suggests that these networks should not exhibit strong characteristics of mentorship (Attribute~\ref{att:mentorship}). We impose a further restriction that the most frequent collaborator within these networks should also be a young researcher (Stages I-III). The age restriction on mentorship is relaxed slightly so that, for instance, we do not require postdocs at the top of the range to collaborate only with somebody younger than them.

To reduce the chances of creating false positives within our initial set of selected researchers, we further restrict the set to those researchers that have published greater than 20 publications in the same calendar year. This profile best represents either high-frequency paper-mill patrons, or foundation authors with a young publication age. This restriction to publication year also imposes a limitation that should be noted as this is artificially excluding a part of the co-authorship graph that may contain fraudulent activity. Some researchers that would otherwise be captured will be missed as their publication peak falls over two years. 

Finally, our model asserts that authorship-for-sale authors should form a loosely connected network (Attribute~\ref{att:authorship-for-sale-netowrk}). To implement this restriction, we take the largest connected graph of researchers for each publication year based on that year's publications and the following year. We have limited the graph to two publication years as we are not interested in measuring the growth of organic research collaboration, but rather the suggested random connections of an authorship-for-sale network. All graph calculations within the paper were undertaken using the python networkX library \cite{Hagberg.2008}. 

\subsection{Validation methods}
\label{sec:2d}
Having identified a yearly cohort of researchers that meet our profile of an ‘authorship-for-sale’ researcher, we then seek to assess how well this correlates with other datasets with high signals of paper-mill activity including the tortured phrases, and the Retraction Watch database. Whilst neither of these sources will account for all suspicious papers, they provide useful indicators to validate our work.

\begin{enumerate}
    \item \textbf{Tortured-phrase dataset.} Tortured phrases \cite{10.48550/arxiv.2210.04895} are awkward English phrases substituted for common language terms (e.g., `flag to commotion' instead of `signal to noise'). While a small portion of these terms may have been legitimately used for non-native English writers, the papers are generally accepted as an indicator of fabricated publication.

A list of papers containing problematic phrases is available from the the Problematic Paper Screener at \href{https://www.irit.fr/~Guillaume.Cabanac/problematic-paper-screener}{https://www.irit.fr/$\sim$Guillaume.Cabanac/problematic-paper-screener}.  Helpfully, each paper in the list is identified by its \textit{Dimensions} publication id allowing an easy comparison based on matching identifiers. 

More than just a dataset, the Problematic Paper Screener represents a new collaborative research integrity investigation methodology.  Issues detected via algorithms (such as tortured phrases), can then be investigated and verified by a volunteer research investigation community, with completed investigations documented on PubPeer. The net result of this workflow is that the tortured-phrase dataset receives significant ongoing review.

As well as tortured phrases, the Problematic Paper Screener dataset contains information on papers that appear to have \textit{clay feet} by citing other problematic papers. Based on our assertion that paper-mill papers are likely to cite other paper-mill papers (Attribute~\ref{att:citation_cartel}), we have also included clay-feet publications in our comparative analysis.

To further assess the effectiveness of using this network signal as an additional method to identify paper-mill activity, we compare the percentage of implicated papers in the tortured-phrase dataset to randomly generated sets of research articles generated in the same time period. 

\item \textbf{Retraction Watch data.} Retraction Watch \cite{q}, a blog series commencing in 2010, has the most comprehensive database of retracted scholarly publications. Within the retraction watch dataset, in 2021 71\% of publications have been retracted due to questionable or fraudulent research or authorship practices. This percentage is also in broad agreement with previous partial studies in 2012, and 2016 \cite{10.1073/pnas.1212247109, 10.1136/bmjopen-2016-012047}

\item \textbf{ORCiD data.}
Finally, we analyse the network for evidence of peer review enablement (Attribute~\ref{att:peerreview}) by identifying researchers within the network with unusually high peer review activity within the ORCiD public dataset \cite{Montenegro.2023}.
\end{enumerate}

\section{Results and Analysis}
\label{sec:3}
In this section, we take the model that emerges from the attributes discussed in earlier sections of this paper and apply it to the graph generated from the data sources described in Section~\ref{sec:2}.  We work through an example in detail and test these results against our validation criteria.

To establish whether unique network shapes (the specific collaboration patterns that emerge from the attributes that describe our model) provide a robust signal for the detection of author-for-sale paper-mill practices, we first establish whether unique network shapes can be identified successfully using the methods suggested (Methods \ref{method:unusualsignal}, \ref{methods:networkshape}). For publication year 2022, Table~\ref{fig:cohorts} breaks down the percentage of researchers by approximate age into unique-network-shape frequency bins. Each bin is an order-of magnitude of frequency. Explicitly, the column labelled ``$> n=0$'' refers to network shapes that occur fewer than 10 times in the whole of the 2022 co-authorship graph. Thus, only 0.10\% of Stage I researchers, and 0.43\% of Stage II researchers have network shapes that have been repeated fewer than 10 times.

\begin{table}
    \centering
    \includegraphics[width=1\linewidth]{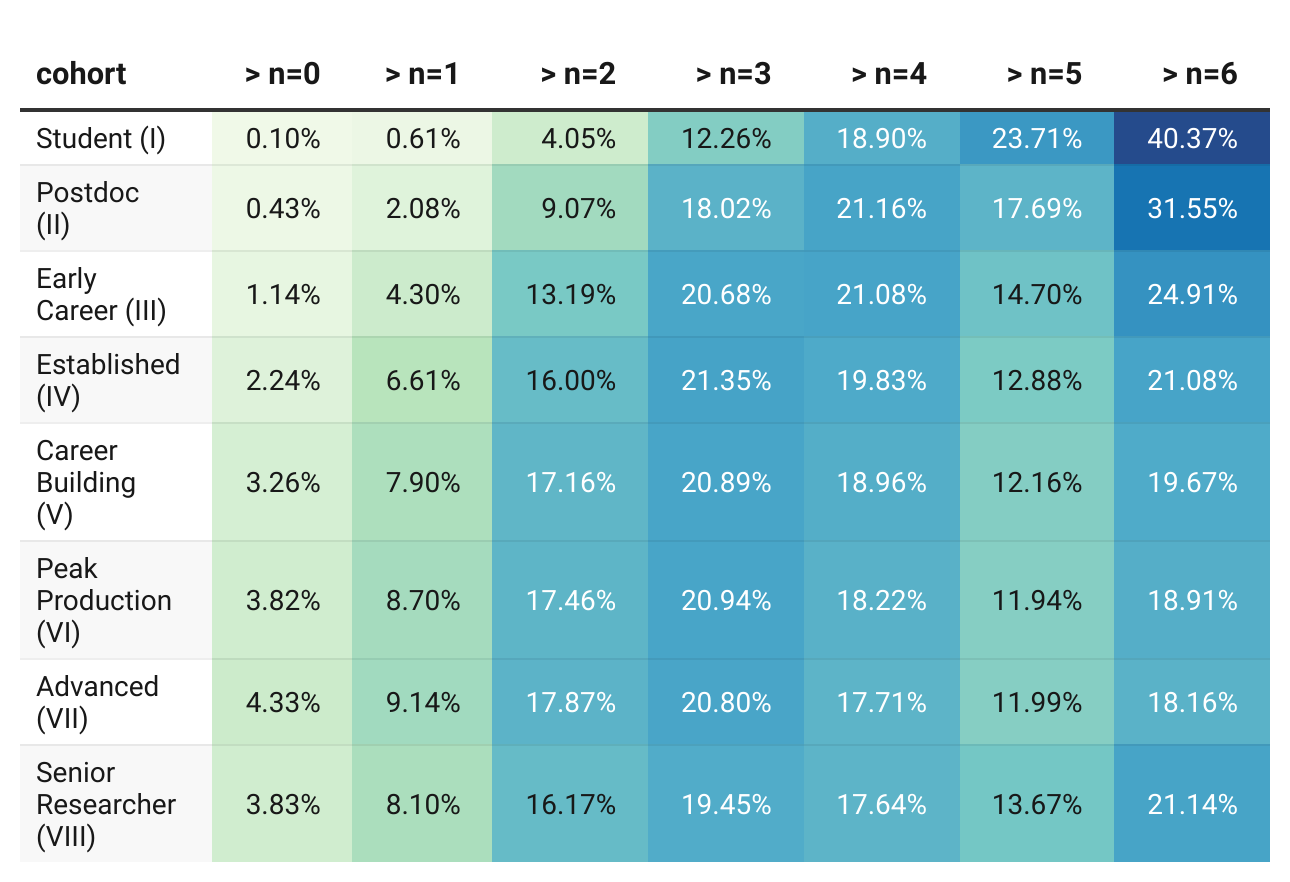}
    \caption{Percentage uniqueness of network shape with approximate career stage for 2022. Each column in the table relates to a specific class of network shape. The column labelled "$> n=0$" are those that have been seen fewer than 10 times; the column labelled "$> n=1$" are those network shapes that have appeared more than 10 times but fewer than 100 times.Percentages are relative to the size of the age cohort. Thus, 0.10\% of students are associated with co-authorship network shapes that occur fewer than 10 times in 2022.}
    \label{fig:cohorts}
\end{table}

\begin{figure}
    \centering
    \includegraphics[width=1\linewidth]{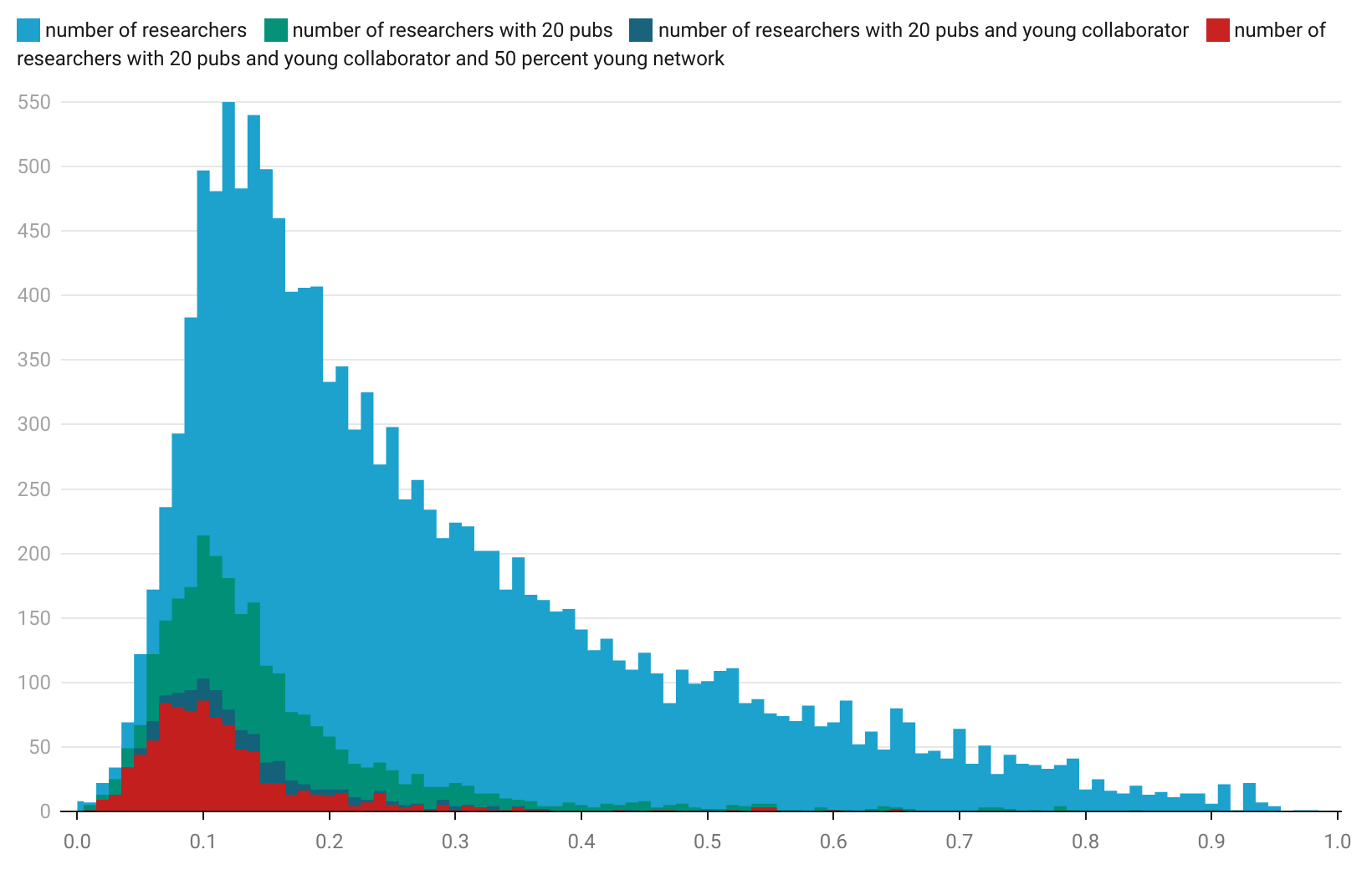}
    \caption{Histogram of Stage I and Stage II researchers with network shapes that have been been repeated fewer than 10 times in 2022. Researchers are distributed across the $x$-axis by their clustering coefficient. The middle green histogram represents the subset of researchers that have produced greater than 20 publications in the year. In the bottom red subset, the most frequent collaborator is a Stage I-III researcher, and greater that 50\% of the network is made up of Stage I-II researchers.}
    \label{fig:cohort_filters}
\end{figure}

Stage I and II researchers are more likely than not to have a network with a low cluster co-efficient (Attirbute~\ref{att:low_clustering_coefficient}), however, there remains a large tail of highly connected networks (see Figure \ref{fig:cohort_filters}).  Figure~\ref{fig:cohort_filters} shows a breakdown of the population of Stage I and II researchers with highly unique collaboration patterns using further attributes.  To be clear, this Figure only depicts data relating to the 0.53\% of researchers identified in the top-left cells of Table~\ref{fig:cohorts}.  The $y$-axis is frequency and the $x$-axis is binned by clustering co-efficient.  The light blue histogram shows the frequency of researchers overall with clustering co-efficient assigned to each division; the green histogram is more restrictive, showing the number of researchers associated with more than 20 publications in 2022; the dark blue (hardly distinguishable from red in shape) show the frequency of researchers with more than 20 papers and their most frequent collaborator is another Stage I-III collaborator on the paper; the red distribution shows the frequency of researchers associated with more than 20 papers in 2022 and where at least 50\% of their immediate collaboration network consists of young (Stage I-II) researchers.

If our hypothesis is correct and that our model detects paper-mill activity, then we can regard Figure~\ref{fig:cohort_filters} as defining bands of probability of likelihood that a researcher is involved in such a network.  The light blue area consists of authors in possession of a highly unique collaboration network.  This may not in-and-of itself be sufficient for someone to be involved in a paper mill - it is merely a statement that their collaboration pattern is irregular for someone of their age, for which there may be many reasons---discipline, geographical situation of university, connectedness of PhD advisor/supervisor, and nature of their specific research project, etc.  However, once we enter the green region we are now focusing on young researchers who have highly unique collaboration networks and who are producing more than 20 papers in a single year. (We remind the reader that high-energy physics papers and the like with hundreds or thousands of authors have been removed from this analysis.)  In order to be productive at this level so early in a career and to have a highly-unique network is suspicious.  

With each additional restriction that we impose, the peak of the frequency distribution moves left-ward toward the region of lower clustering cohesiveness (lower clustering co-efficient), with a heavily suppressed tail at higher levels of connectedness/clustering.  What this means in practical terms is that this the region in which fewer and fewer authors in the network are tightly coupled or come from the same existing networks--put simply--are less likely to actually \textit{know} each other.  If the full context of the situation is taken into account this seems particularly suspicious. The confluence of young researchers without a more senior connector (cf. Attribute~\ref{att:mentorship}), and access to extended disparate networks with high production levels is certainly unusual if not impossible.  
 
 One effect that the criteria applied in Figure~\ref{fig:cohort_filters} are likely to rule out is that these signatures are associated with a new type of brilliant researcher.  When we think of the rise of a talented research we would expect them to be producing results that garnered high levels of attention from serious established researchers, but this is not the behaviour that we're seeing here.  Rather it is a different type of ``brilliance'' - this is a brilliance in which researchers are highly productive in volume but choose to publisher in less recognised and less mainstream venues; one in which they choose by preference to work with younger researchers only; one in which they connect seemingly at random to many different networks and simultaneously make and manage, but then don't maintain, a vast number of research relationships.  At one level this is perhaps a little reminiscent of some social networking behaviour \cite{10.1177/1461444816686104}, but this is not how research tends to work. 
 
 As Dunbar \cite{10.1016/0047-2484(92)90081-j} points out, there is a limit to the number of friendships that we can maintain at different levels of intensity.  Research relationships are not like friendships in that they require significantly more work to maintain and hence, at any one time, one can retain relatively fewer research relationships that form a coherent core to the research on which one is working.  We do not think it likely that this signature suggests a newly emergent class of brilliant researcher, whose characteristic is to be massive connectors of lower level research.  Even if this were to be the case, it might be reasonable to question the value of this type of activity to the research enterprise. The green-shaded region of Figure~\ref{fig:cohort_filters} raises suspicion due to volume of output but we hope that it is clear from the discussion above that the dark blue and red regions highlight behaviour that is highly likely to be worthy of further investigation.

We now turn to a further filter, as outlined in Section~\ref{sec:2} and related to Attribute~\ref{att:authorship-for-sale-netowrk}. This filter makes the assumption that most influential (and detectable) paper-mill networks need to be connected due to the mechanism behind their creation. As a result, we seek the largest connected component of that graph and filter out any small, disconnected components as these are much less likely to be associated with paper-mill activity.

\begin{figure}
    \centering
    \includegraphics[width=1\linewidth]{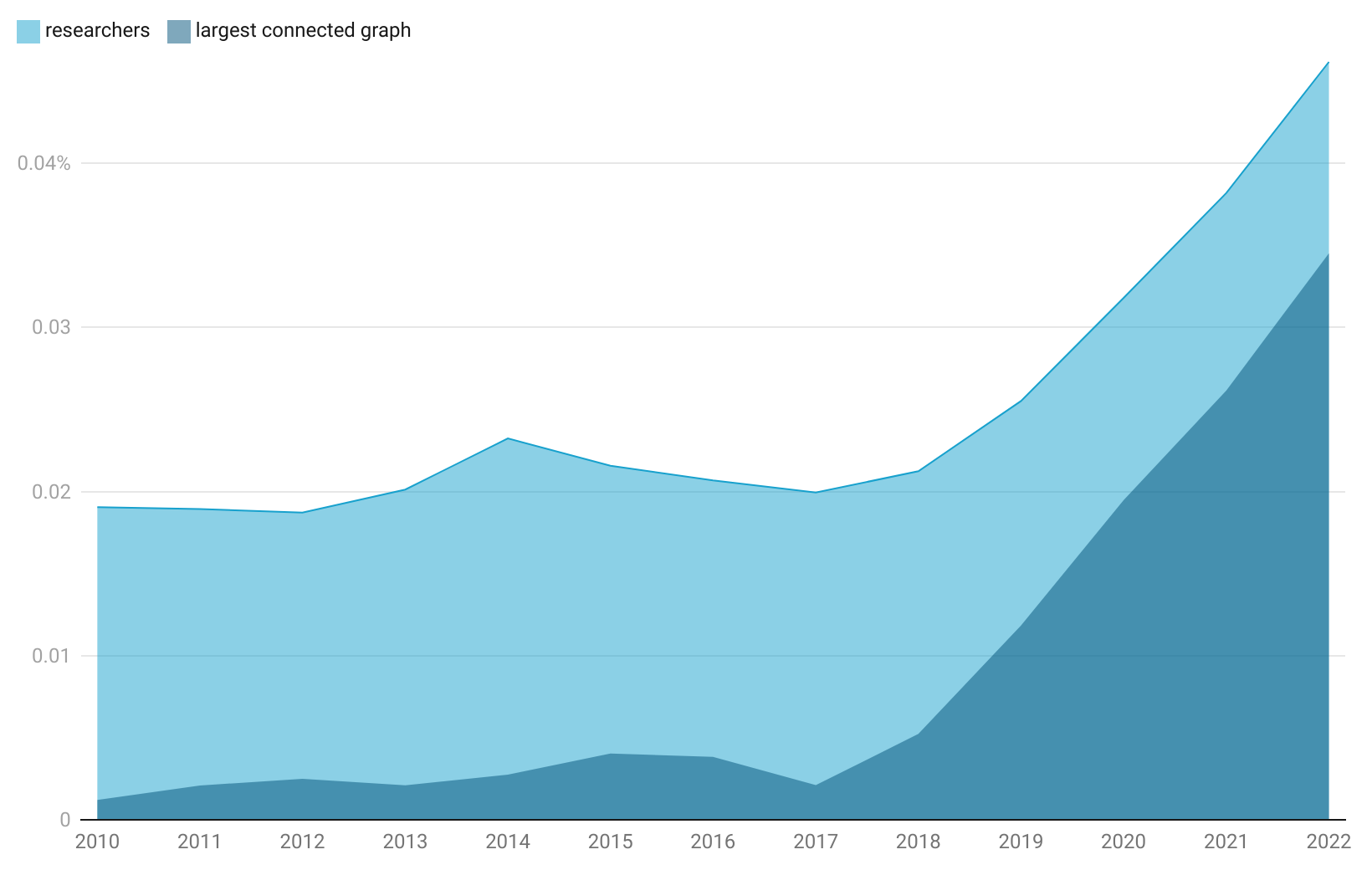}
    \caption{Development of the suspicious author cohort as a percentage of all Stage I and II researchers by year. The two areas are overlapping and not cumulative. The area between the top of the light blue region and the $x$-axis represents all suspicious Stage I and II researchers as a percentage of all Stage I and Stage II researchers.  The dark blue area shows just those that are part of the largest connected component of the suspicious co-author network defined by our model.}
    \label{fig:pubs_researcher_presentages}
\end{figure}

By plotting the signal that derives from the application of these restrictions together in Figure~\ref{fig:pubs_researcher_presentages}, we can observe the relative change in the suspicious author population over time. Overall, the global population of suspicious researchers --- those researchers who meet all our criteria from the discussion of Figure~\ref{fig:cohort_filters} above --- remains small relatively compared with the overall number of Stage I and Stage II researchers (our overall normalisation factor). Prior to 2018,  around 0.020\% of the population could be considered suspicious. From 2018 there is transition in behaviour that doubles the relative occurrence of suspicious researchers over a 4-year period. This appears to be driven by the development of a large, connected network to which a significant proportion of suspicious authors are linked. Indeed, the largest connected component of this group of authors goes from accounting for 0.002\% of authors between 2010 an 2017 (or around 10\% of the suspicious author cohort in that period) to 0.034\% of authors in 2022, meaning that around 75\% of all suspicious authors identified by our model were participants in the largest connected component of the network.

We believe the inflection point that takes place in this plot in 2017 demonstrates clear evidence that there is a coordinating influence at the centre of these activities that started at scale in 2017. From a technical perspective, we argue that the behaviour change in 2017 is termed a \textit{second-order phase transition} in the language of physics and is a behaviour often seen in random graphs dynamics where the number of nodes in the graph is growing with a fixed probability \cite{10.1017/cbo9781316339831}.

\begin{figure}
    \centering
    \includegraphics[width=1\linewidth]{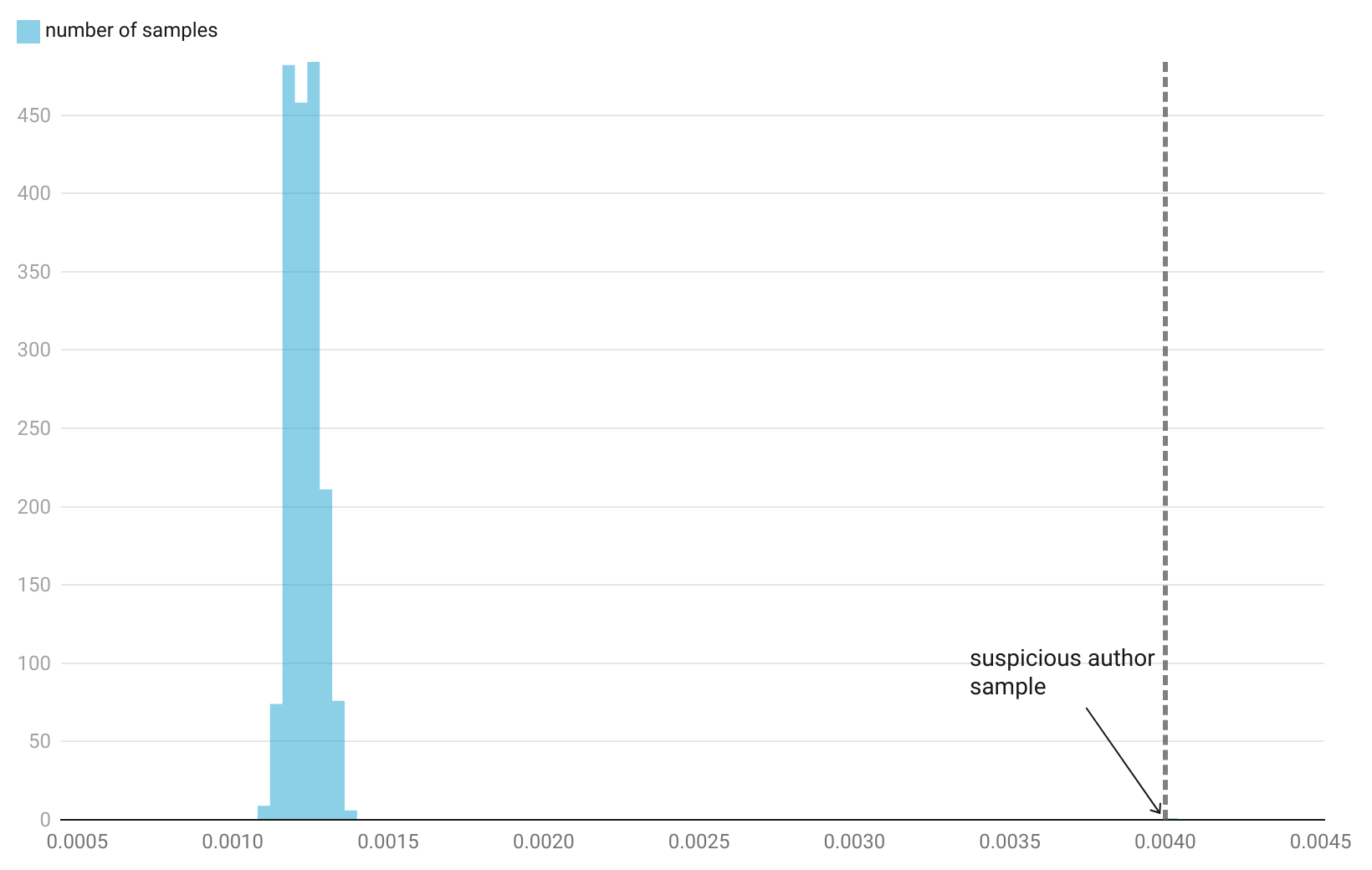}
    \caption{Histogram of the graph density of the co authorship networks of random samples of researchers with clustering coefficients of less than 0.4 and greater than 20 publications.  The sample size of each measurement is the same size as the suspicious author sample }
    \label{fig:density_comparison}
\end{figure}

Although random in nature, there are two important characteristics of the suspicious author graph that make it distinguishable from random connections of non-paper-mill publications. To baseline our results, we studied the network properties of other random samples of the same number of researchers with similar properties to our sample of suspicious researchers. When compared with other equivalent samples of researchers in 2021 (Figure~\ref{fig:density_comparison}),
with a cluster coefficient of greater than 0.4, and publications greater than 20, the graph created by suspicious author cohort had three times the density of the 190 randomly generated samples  (0.004 vs 0.0012 SD 0.000047.) The graph of suspicious authors also differs from the randomly sampled set 
in that the largest connected component of researchers represents a smaller ratio of the total researchers compared, 0.66 compared with 0.80 (SD 0.019) for the sample graphs.  A cluster coefficient of 0.4 was chosen, as most of the sample of the suspicious author set falls below that cutoff
, however the comparison remains unchanged if the cutoff is lowered to 0.1 (N=50, density = 0.0012, largest component ratio = 0.91) or increased to 0.9 (N=210, density = 0.001, largest component ratio = 0.76).

These results are consistent with the authorship-for-sale model expressed above in that there are two different populations of publications that authors are randomly connecting within. In the authorship-for-sale set, the theory would suggest that authors are (mostly) randomly colliding over paper-mill papers. In the equivalent samples, authors are (mostly) colliding over  publications within the organic research graph. As the set of paper-mill papers is (thankfully) smaller than the set of non-paper-mill papers, the `collision' probability of two authors coauthoring on the same paper should be higher, and therefore the graph more dense. This probability is increased further if authorship-for-sale papers have a higher average number of authors (Attribute~ \ref{att:author_numbers}). If the probability of collision is higher, you would also expect the percentage of the largest connected graph to be higher. The fact that it is lower, suggests that there is a number of false positives in the unfiltered suspicious author set. Because these false positives are publishing within the organic research graph, they do not typically collide with researchers that are authors on paper-mill papers. For this reason, filtering on the largest connected component of the suspicious author graph can be seen as an appropriate additional tactic for reducing the number of false positives in the sample. The effectiveness of this approach will be evaluated in the next section.

For the period between 2020-2022, 2056 researchers have been selected as having a profile that meets our suspicious author criteria. It should be stressed that having a profile that matches the criteria of an authorship-for-sale profile does not in itself mean that an individual researcher has participated in research misconduct. How well these profile align in aggregate with other measures and indicators that may imply research misconduct will be assessed in the next section.

\subsection{Assessing the effectiveness of the ‘authorship-for-sale profile’ to identify authors that participate in questionable research}
To assess the veracity of the analysis in Figure~\ref{fig:pubs_researcher_presentages} we would ideally turn to a completely different data source that relies on distinct properties of detection to the model that we have proposed.  In this section we use such a dataset to verify our findings.

The Problematic Paper Screener (which we will refer to as the PPS method or simply `the PPS' for brevity), described in our comments on the tortured-phrase dataset of Section~\ref{sec:2d}, provides an ideal external verification mechanicsm.  The way in which the PPS identifies suspicious authors is purely linguistic and makes use of full text analysis, not relying at all on network structure. Thus, the two methods are complementary and may be used to independently verify each other. In addition, the PPS approach identifies suspicious papers rather than suspicious authors directly, and this point of difference is another aspect of the two approaches that delineates them and strengthens their independence.

\begin{figure}
    \centering
    \includegraphics[width=1\linewidth]{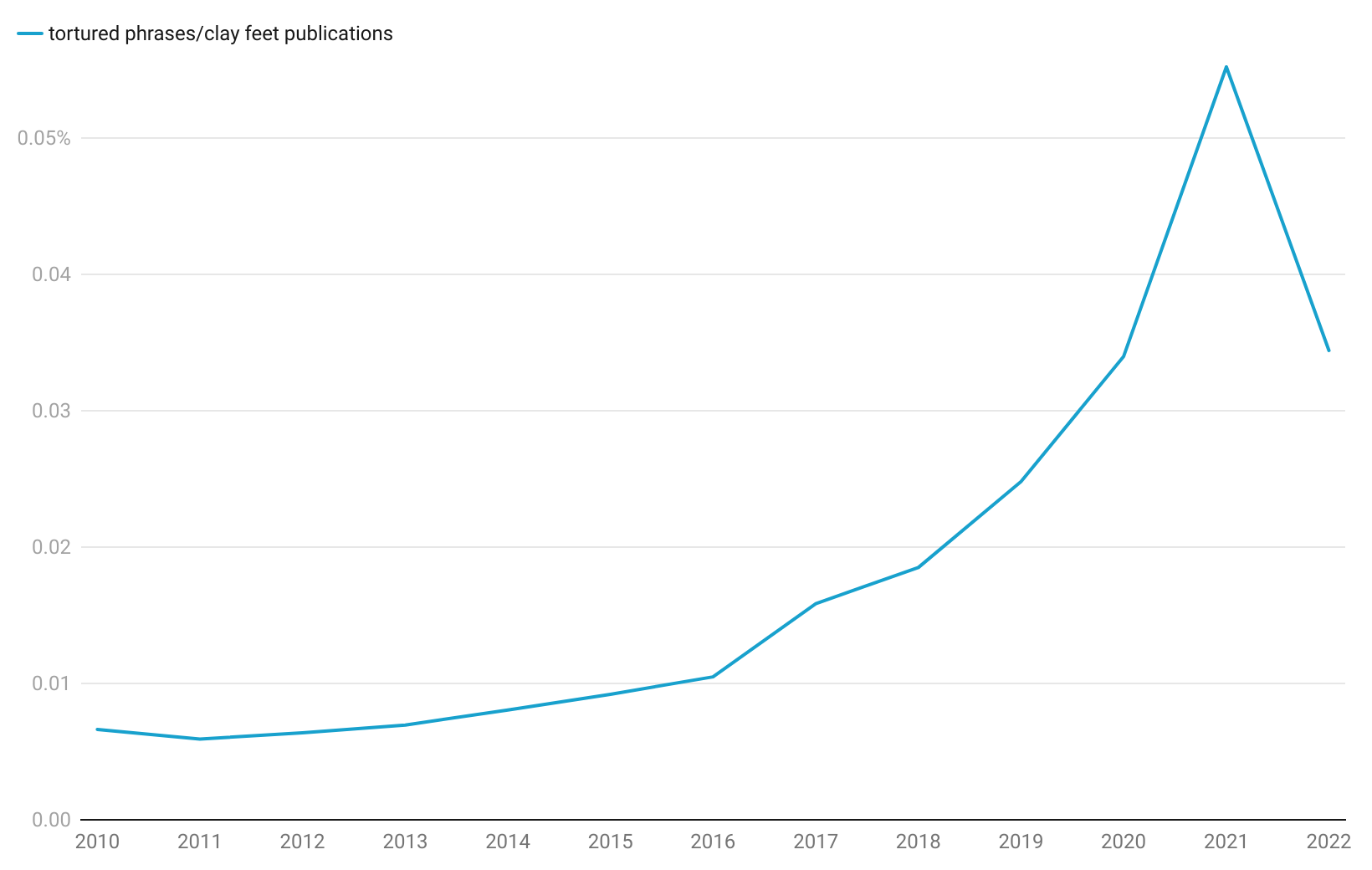}
    \caption{Proportion of papers exhibiting tortured phases as a percentage of all journal articles produced in a year. The tortured-phases dataset was extracted from the Problematic Paper Screener. The drop in papers in 2022 might be explained by the implementation of the tortured phrases detection strategies into submission processing workflows, or a reaction to detection by paper mills.}
    \label{fig:tortured_phrases_graph}
\end{figure}

As an initial test to assess the high-level correlation between the tortured-phrase methodology and the network methodology, we examined the number of research articles identified by the PPS method as a proportion of the total number of journal articles recorded in \textit{Dimensions} each year (see Figure~\ref{fig:tortured_phrases_graph}).  It is at once easy to see the shared features of this figure and our findings shown in Figure~\ref{fig:pubs_researcher_presentages}. 
 The overall shape of Figure~\ref{fig:tortured_phrases_graph} is similar to that of Figure~\ref{fig:pubs_researcher_presentages}, with an inflection point in the PPS method showing in 2016 rather than 2017.  Of course, our data shows the percentage of researchers rather than papers that are suspicious Figure~\ref{fig:tortured_phrases_graph}.  While there are two different quantities, they are clearly correlated up to 2021. As a percentage, the flat region prior to 2016 in the PPS data, is in good agreement with the output reported in our model (the PPS shows between 0.005\% and 0.01\% of total output in this period, compared with 0.002\% to 0.023\% of researchers in our model). The broader range in our model compared with that of PPS makes sense as one model is modelling people and the other is modelling papers.  However, there is a fundamental coupling between the two. The co-authorship graph has papers as edges and people as nodes, but the PPS model effectively reveals the dual of this graph - the graph with papers as nodes and co-authors as edges. In 2022, the percentage of PPS papers that have tortured phrases or clay feet drop away. We assert that this is a real effect and not due to end-of-year effects or incomplete data. Rather it is an example of the dynamic nature of paper mill the interplay between publishers and paper mills as each adopts and adapts to the deployment of new technologies.
 
 The dual relationship between the graphs that underlie this model mean that scaling effects should be correlated (as we observe) but the actual scales of the two graphs and their normalisation is not necessarily fixed. As a result, one can look at variations but one should not read too much into absolute values. It is also interesting to note that the PPS method more quickly identifies paper mills (2016 versus 2017 inflection point), but this too makes sense as our method requires connections to be made in the network (thus mutliple papers need to be published), whereas the PPS method can detect a single paper as soon as the manuscript has been parsed. This speed advantage goes away as soon as the network is established, as any author who has been drawn into the orbit of a suspicious network can instantly be identified. Thus, we should expect the graphs to have a more similar shape as paper-mill practices develop. 
 
To understand whether the similarity between Figures~\ref{fig:pubs_researcher_presentages} and \ref{fig:tortured_phrases_graph} is more than superficial, we need to assess the direct overlap between the papers identified directly by the PPS method and the papers identified through affiliation with identified authors.

To create a tortured-phase dataset for comparison, we focused on publications in the 2020-2022 year range\footnote{Once identified in the Problematic Paper Screener, reviewers are invited to individually assess a paper, and lodge a report in PubPeer  \cite{10.1126/science.349.6252.1036}. In this analysis we have used both assessed and non-assessed publications. Although papers with less than 5 tortured phrases are not shown on the public part of the PPS (to avoid showing false positives), some false positives will remain.}, with a document type of `research article'. Additionally, publications were required to have at least one author that had been resolved to a researcher\_id within \textit{Dimensions}, as papers with no resolved authors would be invisible to our analysis. Of the 3739 publications identified, 10\% (367) included a researcher identified in the ‘authorship-for-sale’ profile, 37\% (1371) included an author that had co-authored with a researcher in the ‘authorship-for-sale’ set. Overall, via this more inclusive measure 72\% of the researchers in the ‘authorship-for-sale’ cohort could be connected back to the tortured-phrase dataset, compared with 16\% who were direct authors. 

There was no difference in the proportion of tortured-phrase or clay-feet publications between the matched and unmatched sets, however there were some differences in authorship patterns. Unsurprisingly, papers with fewer authors were less likely to be matched back to the ‘authorship-for-sale’ network. Articles in the unmatched set contained an average of 3.7 authors, compared with 5.0 authors for the matched set. This difference is increased when authors that can be matched back to a unique researcher ID are taken into account (2.6 compared with 4.4 authors). Publications within the matched set were also far more likely to have international co-authors (50\%) vs 8\% for the unmatched set. Of the unmatched set, 56\% were also a single institution, vs 27\% for the matched set. The bias towards a greater number of authors as well as greater percentage of international co-authors aligns well with the expected behaviour of authorship-for-sale networks (Attribute \ref{att:author_numbers}).

\begin{table}
    \centering
    \includegraphics[width=1\linewidth]{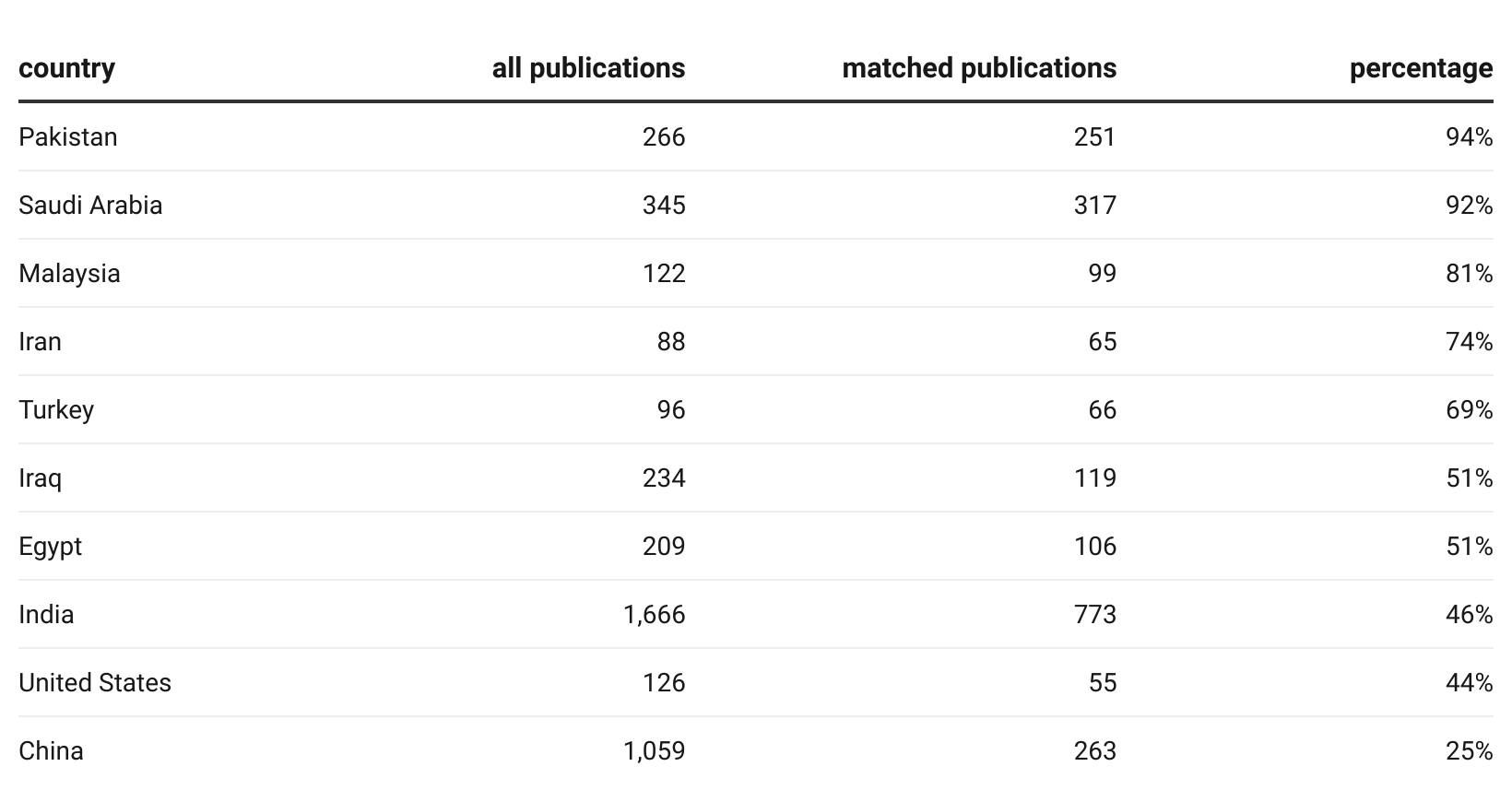}
    \caption{Absolute number of suspicious papers detected in the years 2020, 2021 and 2022 by country. The ``All publications'' column is the total number of publications identified via the Cabanac method, the ``Matched publications'' column reports the number of publications matched with the model proposed in the current paper.  The ``Percentage'' column gives the proportion of matched publications.}
    \label{fig:tortured_phrases_network_country}
\end{table}

A further test that we conducted aggregated suspicious papers by a variety of different axes to search for implicit biases in the data (or, put another way, to see whether the datasets share similar findings). Upon examining geographical aggregation we find that the profile of matched and unmatched articles also differed by country of author affiliation. For the largest two country cohorts in the top 10, only 25\% of articles with Chinese authors were matched, with 46\% for India. The highest rates were for Saudi Arabia, and Pakistan (94\% and 93\%) (Table~\ref{fig:tortured_phrases_network_country}). This difference reflects the skew in the countries of authors identified by the authorship-for-sale profile, however given the the greater prevalence of domestic and single institution papers in the unmatched set, it may also reflect a split in the way paper-mill papers are constructed, with one model focusing on selling entire papers, and another focusing on selling authorship places. 

The data in Table~\ref{fig:tortured_phrases_network_country} suggests that there is a good correlation between the tortured-phrase and suspicious author datasets, however, a final step is required to show that these overlap percentages are significant.  We must test against a null model---specifically, we want to see if the model is susceptible to detection of false positives. In other words, can this method discriminate between a suspicious paper signal and background publication noise?   To do this, we use a numerical Monte Carlo simulation---we randomly selected groups of publications of publications from Dimensions.  These groups were designed to be of the same size and from the same time period as the tortured phrases dataset. We then compared the incidence of tortured papers and model-identified papers between these random sets. Authors for the ‘authorship-for-sale’ profile appeared 1.2\% of the time (SD = 0.15).  Authors that had previously collaborated with one of the authors in the authorship-for-sale profile appeared on 16\% of papers (SD 0.63). These rates are significantly lower than the suspicious author overlap figures of 10\%  (direct) and 37\% (collaborated) for the tortured-phrase dataset, indicating that our method for identifying papers belonging to suspicious authors is effective at identifying paper-mill content without also identifying a high number of false positives.



All of the analyses above pertained to the highly restricted dataset, where we filtered to the largest connected component. We extended our analysis further to examine the overlap between the PPS method and the set of papers identified by association with the less restricted group of suspicious authors including those who are not (yet) part of the most connected subgraph.  When compared with the tortured-phrase dataset, 1\% of the potential suspicious authors outside of the largest connected graph matched to the tortured-phrase dataset, and 20\% have coauthored with a author on a tortured-phrase dataset. These rates are significantly lower than the rates for the connected graph (16\% and 72\%).  As foreshadowed, the much lower match rates justifies removing these researchers outside of the largest connected component of the network from the suspicious author set.

A further sense check that was carried out (as described in Section~\ref{sec:2} was to perform a similar comparison with the Retraction Watch publications as was performed with the tortured-phrase dataset described above. Articles within the Retraction Watch database were filtered to include only those with a retraction reason associated with a breach of research integrity \cite{10.6084/m9.figshare.24948789}. Overall, authors identified in the suspicious author profile appeared on 7.43\% of the 1858 research articles articles listed in the Retraction Watch database between 2020 and 2022, over six times the rate established above for randomly selected sets of publications. Although this percentage is lower than the 10\% overlap with the tortured-phrase dataset, this is to be expected as retractions cover inside-the-tent issues of research integrity as well as those that arise from paper mills. 

\subsection{Proximity to high-volume peer reviewers}

In order to facilitate publications ending up in reputable journals, paper mills must rely on complicit peer reviewers to push publishing decisions through. Although it is not possible to directly identify complicit peer reviewers, it is possible to use the public ORCiD data file to identify peer reviewers with an unusually high number of peer reviews. ORCiD records are matched back to \textit{Dimensions} researcher records, allowing publication age, and connection to the suspicious author co-authorship graph to be calculated.

Between 2020 and 2022 (in keeping with our previous analysis), we were able to identify 741 Stage I or II researchers that had recorded greater than 250 peer reviews over three years. Of these, 8.2\% of researchers were also part of the ‘authorship-for-sale’ cohort, and 51\% had previously co-authored with someone in the suspicious author cohort, representing 218,907 claimed peer reviews. Of the 49\% of peer reviewers not matched to the co-authored cohort, 143,066 peer reviews were recorded.  The high degree of connectivity between these high volume peer reviewers and ‘authorship-for-sale’ cohort, provides a new perspective on the observation that peer reviewers are more likely to be favourable to authors within their network \cite{10.1016/j.joi.2019.03.018}. Some allowance for double counting must be made, as in rare cases reviews for the same journal have been reported in ORCiD from both Publons \cite{10.1038/nature.2014.16102} and the publisher directly. A breakdown of high-volume peer reviewers by country shows a similar distribution to the tortured-phrase dataset (Figure~\ref{fig:orcid_country}) although Saudi Arabia is less dominant.

Without direct access to publisher records, it is tempting to make a link between the 218,907 peer reviews and the 84,075 publications authored by individuals in the suspicious authors cohort. However, the overlap is less apparent at the journal level. If we define a review ratio as the total number of reviews divided by the total number of publications, then for the 181 journals producing more than 50 publications per year, there is one review for every five publications on average, and only 11 journals have a peer review to review ratio of greater than 1 in 2. Nonetheless, the co-authorship proximity of a high-volume peer review network is notable - especially given that these reviews are just the ones that have been made public.

\begin{table}
    \centering
    \includegraphics[width=1\linewidth]{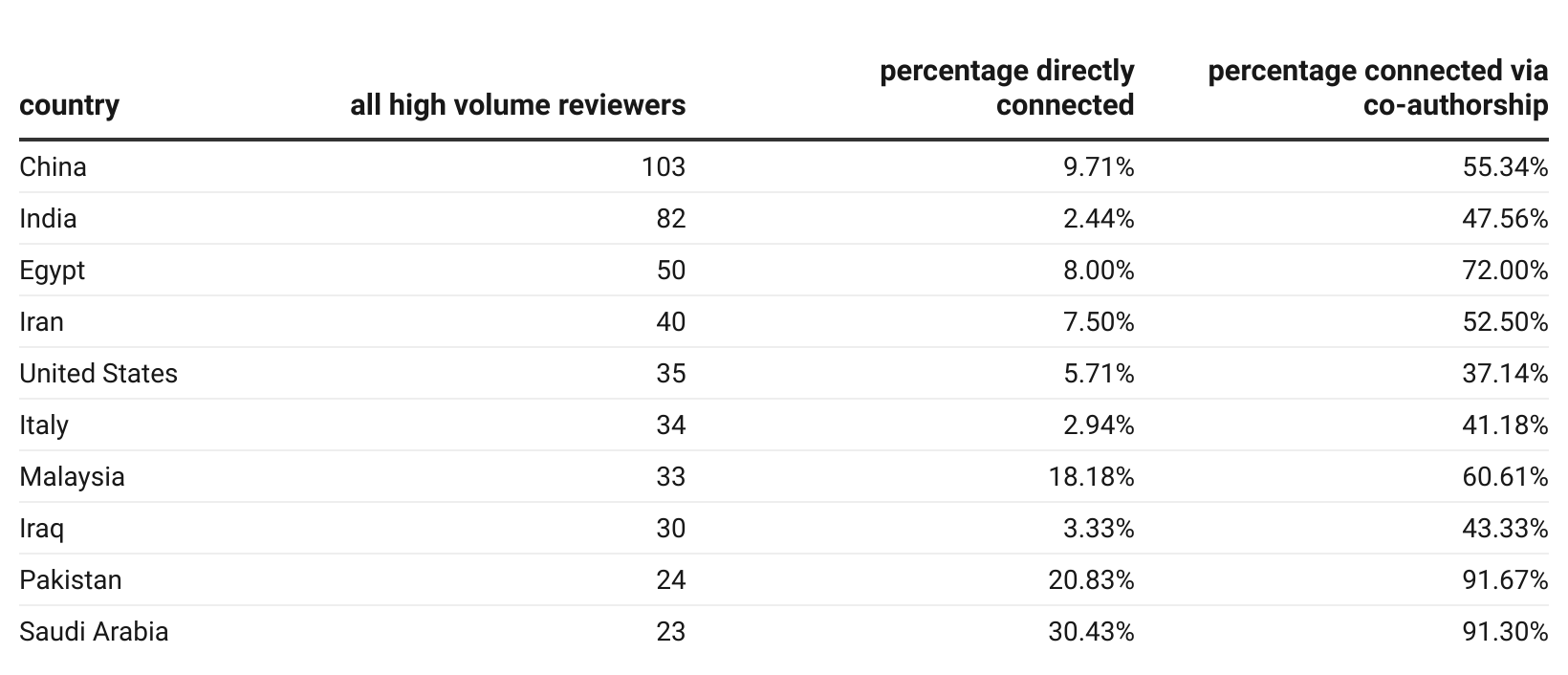}
    \caption{Number of researchers with ORCiD ids that have more than 250 peer reviews in the years 2020-2022 by country of their most recent affiliation. The percentage overlap to researchers in the suspicious author cohort either directly or via co-authorship is also shown.}
    \label{fig:orcid_country}
\end{table}

\subsection{Applications}

Based on the analyses above, we have shown that it is possible to identify young researchers having characteristics that are shared with authors appearing to have frequently participated in authorship-for-sale transactions - either as high-frequency customers or as foundation authors. These researchers are young by publication age (approximately Students or Postdocs), have an egocentric networks with a low clustering coefficient, and yet are connected to each other. When taken together, this cohort aligns well directly or via one degree of collaboration with the tortured-phrase dataset. The network also has access to a large number of high-volume Peer Reviewers. 

Further, we have shown that direct detection rates in random sets of papers are significantly lower than the detection rates in the tortured-phrase dataset. This means that detecting the presence of an elevated `suspicious author' signal in a collection of publications should be a cause for concern. In this section, we explore the application of this measure to publishers, journals, national systems, and institutions.

\subsubsection{Identifying publisher risk profiles}

The initial motivation for this paper was the observation that paper-mill papers with more convincing author profiles were more likely to pass through the submissions and review process. If the ability to flag authors as belonging to a suspicious author network were implemented at the beginning of the submission process, how many papers would require additional review?

\begin{figure}
    \centering
    \includegraphics[width=1\linewidth]{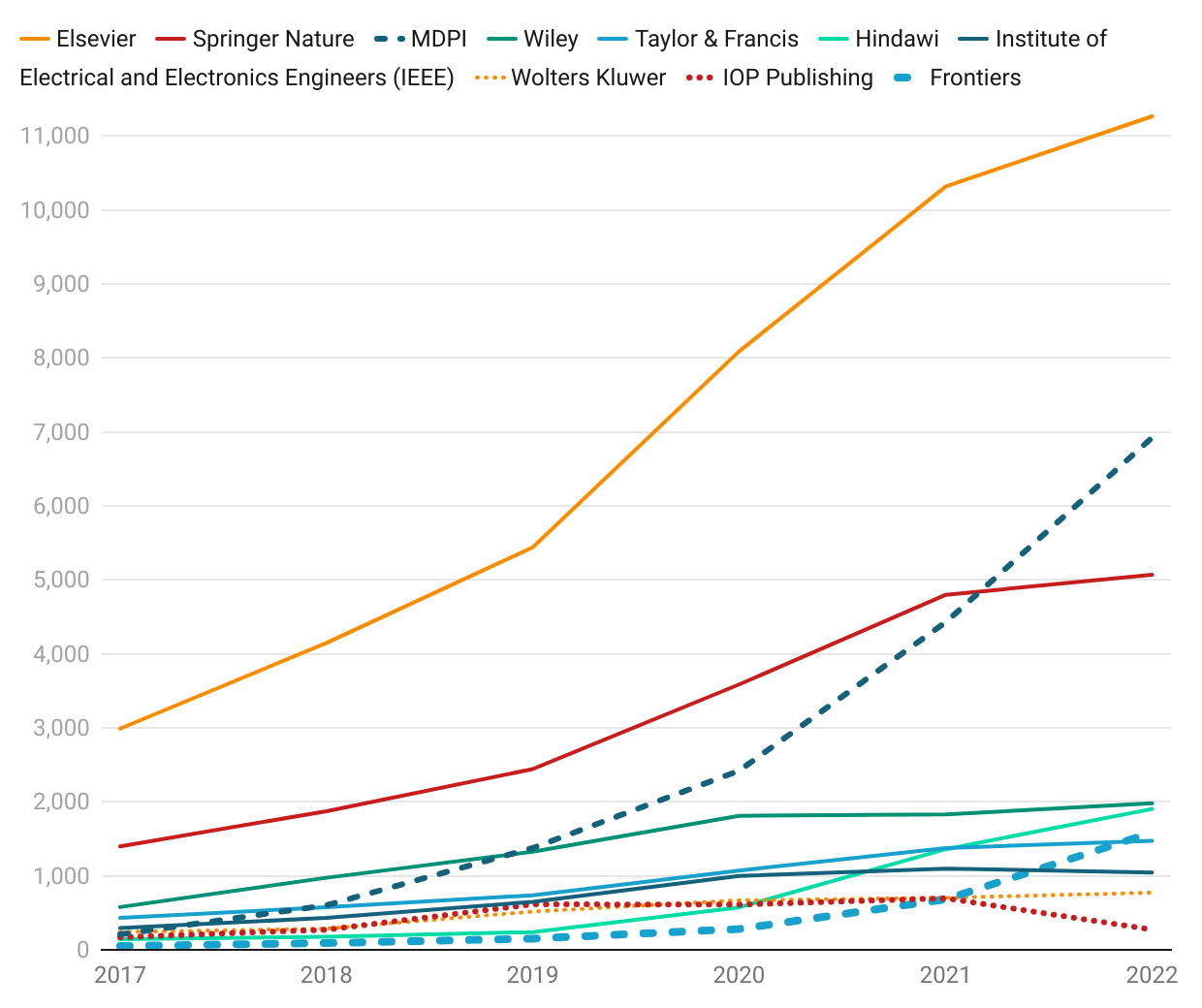}
    \caption{Publications by volume that have an author in the suspicious author set}
    \label{fig:publisher_risk_profile_raw}
\end{figure}

Figure \ref{fig:publisher_risk_profile_raw} depicts the number of publications by publisher that involve an author from the suspicious author cohort (2017-2022) and identifies the number of papers that require close review. By 2022, Elsevier had published more than 11,000 implicated publications, with MDPI just under 7000, and Springer Nature at over 5000 publications. If review is only handled at the publication level, all of these publishers would have a significant amount of work to review these publications. Given that there are only in the order of 1700 researchers associated with these papers, and further, that these authors would be common across publishers, a joint investigation strategy that focused on authors rather than individual publications is likely to more efficient.  

\begin{figure}
    \centering
    \includegraphics[width=1\linewidth]{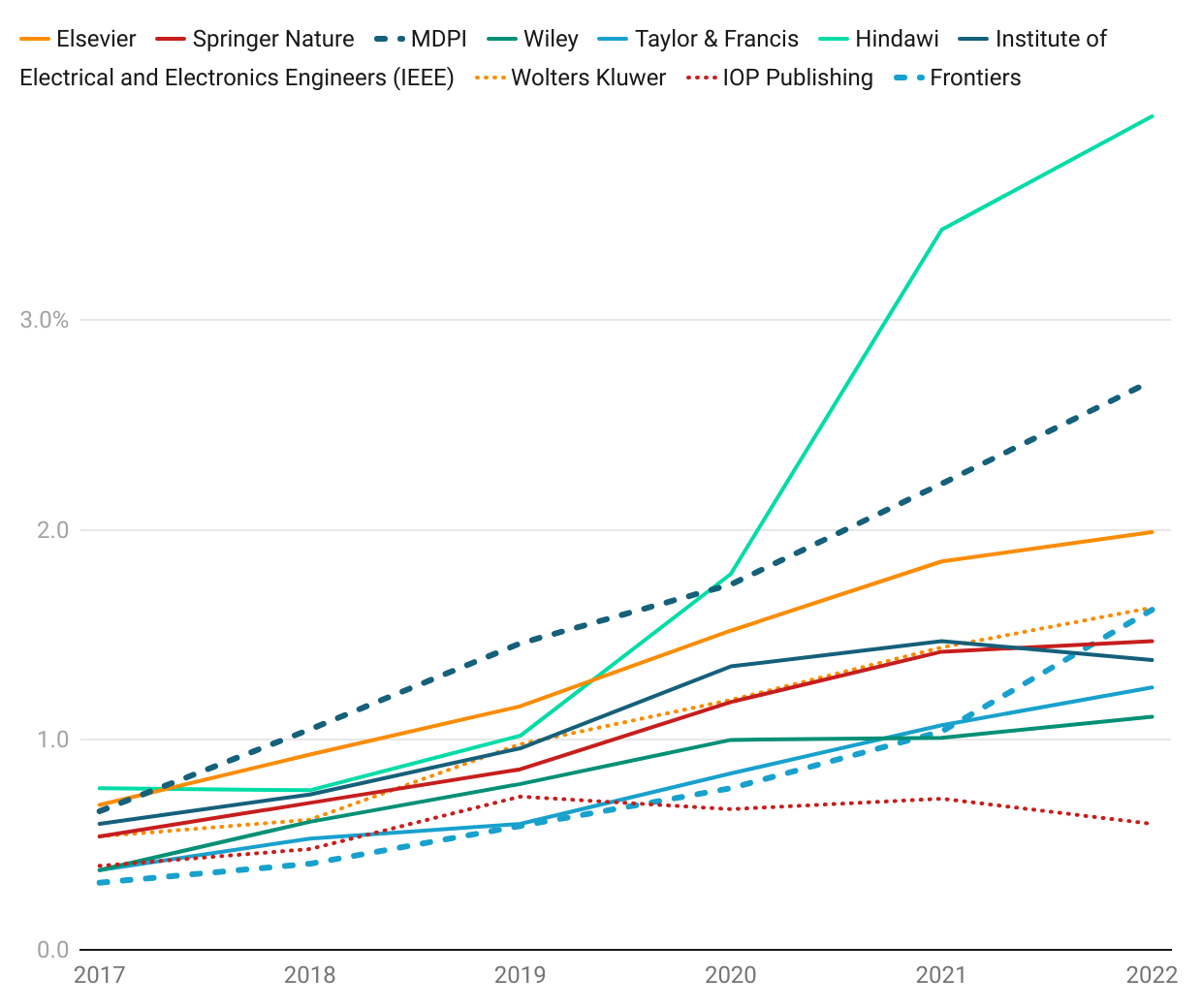}
    \caption{Risk profile of exposure to the suspicious author cohort. The percentage of publications exposed is expressed as a percentage of publisher output.}
    \label{fig:publisher_risk_profiles_percentage}
\end{figure}

Aside from identifying the individual papers that need to be investigated, a comparative risk profile can be calculated for each publisher based on the overall network statistics associated to publishers via the papers that they accept. This can also be thought of as a capability profile to assess the abilities of publishers to detect fraud and to safeguard the scholarly record against paper-mill activity.  As Figure~\ref{fig:publisher_risk_profiles_percentage} highlights, our method appears to suggest that Hindawi began to assume a significantly higher risk profile from 2020 onwards. By 2022, our model indicates that Hindawi had doubled its risk profile expressed as the percentage of implicated publications to 4\%, with MDPI steadily climbing to just under 3\%, and most other publishers sustaining a risk profile of somewhere between 1 and 2\%. Whilst the recent struggles with paper mills at Hindawi have been well documented \cite{40i,ao},  it should be sobering to note that Hindawi has only double the risk profile of the majority of other publishers, according to our model, and that risk profiles of implicated papers can change rapidly over a small time period. Moreover, whilst the absolute number of implicated papers is staggering, the percentage of implicated publications in the scientific record for most publishers is lower, but not much lower than the reported minimum 2\% paper-mill submission rates at the journal level \cite{p6}. For the highest volume publishers, it could also be argued that a higher standard of probity should be required as they are able to have a disproportionate effect on the viability of paper mill business model.

\subsubsection{Identification of journals of concern}
\begin{figure}
    \centering
    \includegraphics[width=1\linewidth]{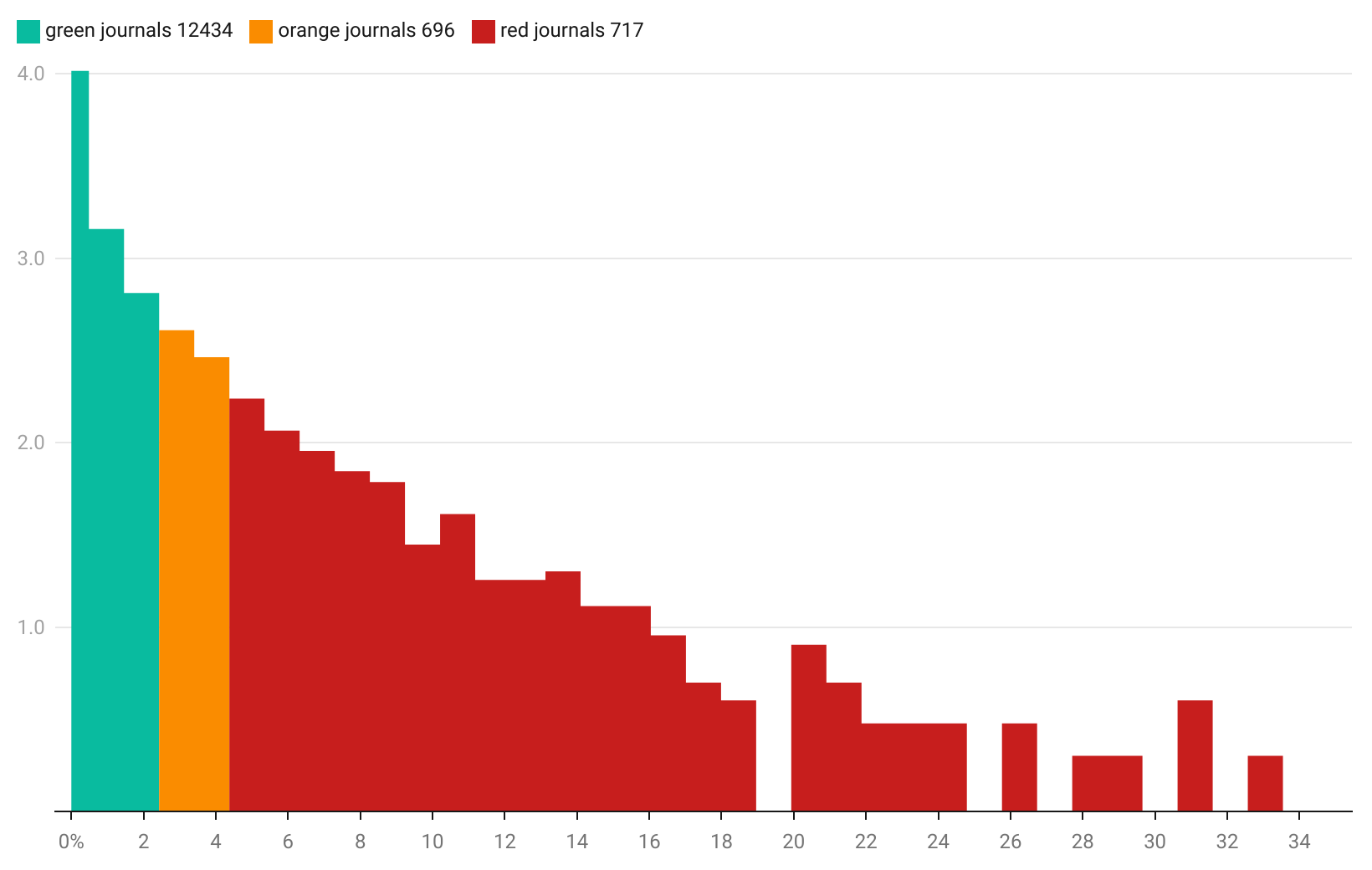}
    \caption{Histogram of journal exposure suspicious author cohort.  The $x$-axis is the percentage of papers in any given journal that appear to be associated with the suspicious paper network (i.e. number of papers found in a journal that appear to be authored by a suspicious author).  The $y$-axis, using a $\log_{10}$-scale, is the frequency with which journals fall into each 1\% band of exposure. There are 12,434 journals (shown in green) that contain potentially suspicious-author-affiliated publications at a rate between 0-2\%; 696 journals (shown in orange) have an exposure of between 2\% and 4\%  of their articles; 717 journals have an exposure to the suspicious-author network with more than 4\% of their articles.  Note that the $log_{10}$ scale of the $y$-axis compresses and underplays the number of ``safe'' journals and tends to emphasize the size of the tale of this distribution.}
    \label{fig:hist_journals_2022}
\end{figure}

As reported in the COPE and STM report on paper mills, paper mills target some journals more than others, especially when they have been successful in the past. At a more granular level, the suspicious author profiles identified in this paper can be used to identify specific journals of concern. Figure \ref{fig:hist_journals_2022} provides a log scale histogram of journals and their associated risk percentages. Only journals with greater than 50 publications in a year have been assessed. Of these 12,434 journals (90\%) have little association  (<2\% of papers) with authors identified in the `authorship-for-sale' profile. Just over 10,000 journals have no direct overlap with the researchers identified in this paper at all.

Noting that 4\% is overall rate for Hindawi, and the overall rate for most publishers is between 1 and 2\%, across the publishing industry there are 696 journals with concerning percentages of implicated papers between 2 and 4\%, and 717 journals with highly concerning risk profiles at greater than 4\%. Armed with this analysis, publishers should be able to target their analysis to specific high risk journals. 

\subsubsection{Profiling the selected researchers by country}

Like publishers, the reputation of national research systems and institutions can also be harmed through the association of their affiliation with paper-mill content. 

Table \ref{tab:country-institution} provides a breakdown by country and institution. What is immediately apparent is the Saudi Arabia is over represented when compared with the size of its research population. Of Saudi Arabia's Stage I-II workforce, 1.2\%  have been identified as having a suspicious author profile, with 22.6\% percent of research articles and reviews implicated. Whilst this does not necessarily mean that 22.6\% of Saudi Arabia's publications have been produced by paper mills, it does suggest that there is strong reason to ask why these numbers differ so significantly from other countries. 

\begin{table}
    \centering
    \includegraphics[width=1\linewidth]{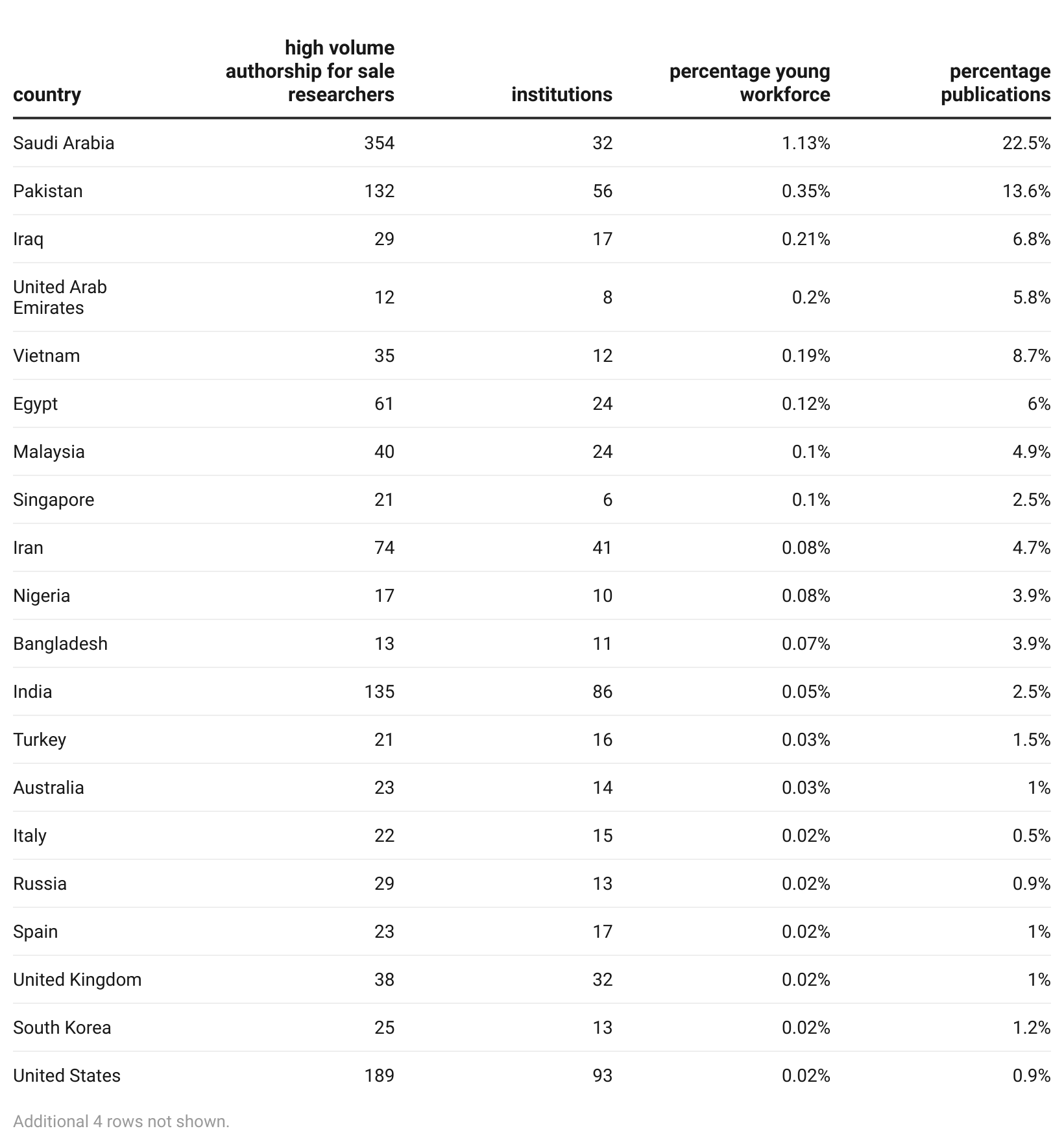}
    \caption{The number of researchers and institutions by country within the suspicious author set.}
    \label{tab:country-institution}
\end{table}

\subsubsection{Institutional reputation protection}

At an institutional level, being associated with a large number of paper-mill papers that have not been retracted would reflect poorly on institutional culture and research integrity training. On the other hand, evidence of proactive activity in retracting papers on behalf of institutional authors reflects well on an institution \cite{10.1371/journal.pmed.1001563}. 

In the publisher use cases above, we have focused on identifying papers where an author in the suspicious author cohort is specifically an author on a paper. This is because using broader measure of connected authors (15\% of papers in the random set) is too high to be useful. At an institutional level however, connected authors can be used to assess the exposure of institutional researchers to the suspicious authors network.  

Where institutional researcher connections to the suspicious author network are found, interventions need not be time consuming. For researchers where the exposure is through a single paper, the action required by the institution might just be to make the researcher aware of the connection, and ensure that they were aware of the authors involvement in the paper. For senior researchers with a large amount of exposure, a more formal investigation may be warranted. For less senior researchers, who are often the targets of authorship-for-sale enterprises, the action might require engaging in a conversation with their supervisor to understand how they came to be involved in the paper, and supporting the researcher in initiating a request to retract the paper. Interventions such as these are invaluable as they are unlikely to be initiated by the junior researcher, and remove the possibility that the researcher can be coerced later in their career into further participation in the authorship-for-sale enterprises. Further, if institutions are able to identify and remove `foundation authors' (Attribute~\ref{att:foundation_authors}) that are operating under their affiliation, then institutions can play a pivotal role removing the ability of paper mills to construct convincing author networks in the first place -provided the work required by individual institutions is not overly onerous.

Table \ref{tab:institutional_review} provides a country level view of the average number of researchers per institution that have some exposure to the `suspicious author cohort.' Only institutions with greater than 3000 unique researchers in 2022 based on publication affiliation have been considered here. For most countries, the average number of researchers, and therefore interventions required, falls between 20 to 80. For countries that are more implicated in the suspicious author network (such as India  with an average number of 147 reviews per institution and Iran with 296), the average number of interventions are higher, but still manageable. Alongside the main figures, the number of exposed junior researchers is also provided, giving an indication of the amount of interventions where extra effort is required. 

\begin{table}
    \centering
    \includegraphics[width=1\linewidth]{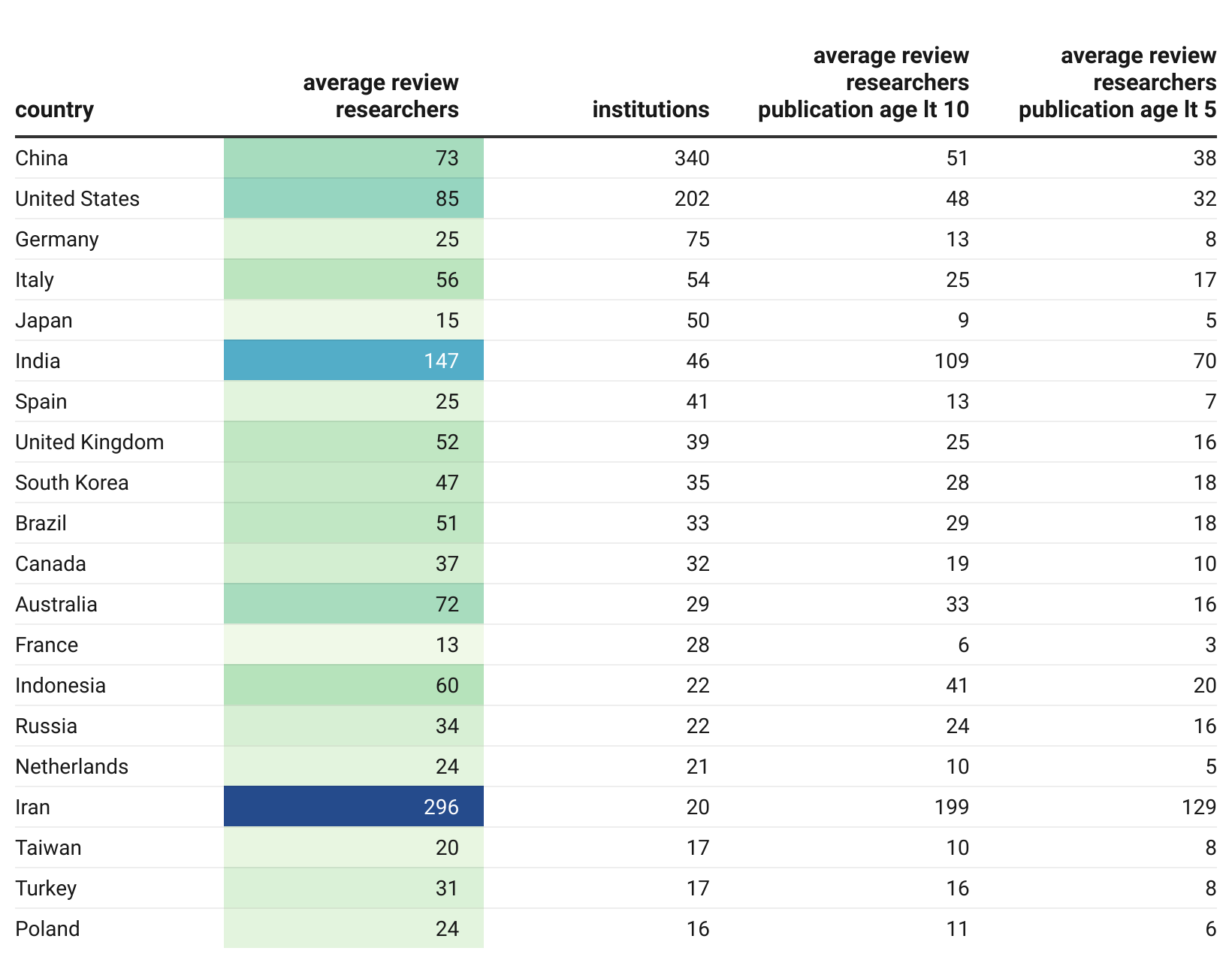}
    \caption{Risk profiles by country and institution of exposure to the suspicious author set. Only institutions with more than 3000 researchers in 2022 are shown.  For each country the average number of reviews required per institution is displayed, along with the number of institutions that this effects. The average number of reviews required is also further analysed to show the average number of younger researchers (by publication age)  that these reviews would involve. }
    \label{tab:institutional_review}
\end{table}

\subsubsection{A collaborative response to the problem of paper mills}

Through the use of the authorship-for-sale co-authorship network, institutions and publishers have an opportunity to collaborate on removing paper-mill content. Institutions benefit from having a relatively low number of researchers to investigate, with established lines of communication to researchers who were most likely not the corresponding author. If pursued proactively, institutions can also play a pivotal role in cutting off the supply of convincing author profiles that can be included in paper-mill papers.

Publishers for their part have a new process available for identifying and preventing paper-mill content from being published. It is hoped that over time, this could also lead to a decreased amount of investigative work required at the institutional level.   

\section{Discussion}
\label{sec:4}
We have demonstrated an approach to identifying paper mills that complements existing methodologies. In our method, rather than focusing on analysis of the text of the manuscript as established methods do, we focus on identifying irregularities in the context of the article.  We have shown that it is possible to develop a model that describes some facets of paper-mill activity and have shown this model to yield results that are comparable to those of the PPS method.  We have also shown that this method discriminates effectively between background noise and a signal associated with paper-mill operation.  It is noteworthy that the model described in this paper is only one model that can be developed to detect paper-mill signals in contextual data---our approach communicated here focuses on a specific collection of facets based on a particular set of assumptions for how paper mills operate---more research and a better understanding of paper-mill operation could lead to a set of models or an extended model that could become a more powerful tool.

Established technologies that detect suspicious behaviour, as pointed out above, use text mining and other computationally expensive approaches.  Thus, while these are excellent methods to understand the structure of a manuscript, they do not necessarily scale well and, with improving technology, constantly need to be developed.  Our model has different scalability characteristics, which is helpful as a complement to existing strategies.  Once the the network of suspicious authors is calculated, then those connecting to that network are instantly worthy of further investigation.  Consequently, our approach has the facet that the network calculation needs to be carried out frequently but it is done for all papers that need to be checked rather than being done individually per paper.  Calculations of the nature required for this purpose have not been easy to access until the development of cloud compute capacities and its application in scientometrics and bibliometrics \cite{10.3389/frma.2021.656233}.  Thus, we believe that a strength of this method is that it is both may efficient and cheap to use at implement scale, and less susceptible to adaptation than than methods that focus on the technology of paper mill production. However, like the PPS methods described above, our method is only identifying outliers with a statistical probability and, as such, is no silver bullet.  Rather, we believe that it serves as a useful method to enhance and extend methods such as the PPS to improve confidence levels and to help to detect a wider class of papers that may be suspicious.

Of course, our method is not without significant challenges and it is important to be aware of the weaknesses of the approach that we have described in this paper.  In the remainder of this section we reflect on the limitations of our approach and identify several opportunities for the application of our approach. Finally, we make some suggestions for future avenues of exploration.

\subsection{Limitations}

\begin{enumerate}[label=(\alph*)]
\item \textbf{Author disambiguation.} One limitation of analysing the authorship-for-sale network is that a focus on young researchers is a region of name disambiguation that is the hardest; there is the least amount of publication information over which to base an identity decision. Author disambiguation is also more challenging for some countries with written character systems \cite{10.1177/01655515211018171}. Even in countries where author disambiguation is easier, a number of factors will mean that false positives are unavoidable. Even state-of-the-art algorithmic author disambiguation is fallible, and will sometimes mean that two (or more) distinct authors are either merged into the same profile or a single author has multiple profiles. Both of these issues with automated disambiguation cause problems for the confidence that one can have in our methodology.  In the first case, an author may be flagged as suspicious when they merely have a similar name to another author who actually \textit{is} suspicious---i.e. potential for maligning an author.  In the second case, an author may have multiple unique identifiers and be able to evade detection as only a subset of their work is associated with the suspicious author network.  Indeed, this latter use case is the basis for a strategy to evade our method.  For these reasons, it is important that all authors identified by the methodology outlined above have humans curate the findings before any action is taken. Further, there must be a process to disentangle researcher identities when disambiguation errors are detected.

In this context, it is clear that ORCiD adoption is an important tool in combating paper mills.  Requiring authors to register an ORCiD and to associate their ID with the paper at the time of submission decreases reliance on fully automated disambiguation methodologies and strengths our approach.  Relatively low ORCiD adoption rates \cite{10.3389/frma.2022.779097,10.1002/leap.1451} are effectively keeping a low bar for paper-mill activity. The call for system wide change with regard to ORCiD is not new \cite{10.1073/pnas.1715374115}, but it is certainly now more urgent.

We also realise that this recommendation does not solve all problems---ORCiD identifiers are being misused or registered afresh for each new paper-mill paper \cite{10.1007/s11192-021-03996-x}.  However, the date of ORCiD registration is easily measured, and a set of new ORCiD records, alongside authors with little history raises its own red flag. Having an a ORCiD is not on its own a sign of upstanding research integrity, but a journal that enforces their use across all authors makes the investigation of research integrity issues far easier to manage.

Beyond an investigation into paper-mill authorship, this study has also shown that the use of ORCiD's in recognising peer review is also useful as a tool in identifying researchers for whom---by volume of peer reviews alone---should not be requested to review further papers. This signal is immensely valuable as it covers all publishers. Where acceptable, we strongly suggest that all publishers mandate the registration of peer reviews. Removing the network of peer reviewers that enable the publishing of authorship-for-sale papers is equally as important as dismantling the social network of authorship-for-sale authors itself.

\item \textbf{A limit to `outside-the-tent' research misconduct.}  As identified in the introduction, a further limit on taking a research network approach to the identification of suspected paper-mill activity is that it is only effective in identifying behaviour where researchers have had to go `outside-the-tent' of their local research network to participate in paper-mill networks. `Inside-the-tent' misconduct such as image manipulation, plagiarism, data fabrication, or even the purchasing of entire research papers to populate with local authors will not be detected. For these papers, research misconduct detection techniques that focus on the content of the paper remain the most effective approach. 
\end{enumerate}

\subsection{Reflections on data driven research integrity collaboration between institutions and publishers}

By basing the identification of issues concerning research integrity on externally accessible linked data sources, opportunities arise for collaboration across multiple publishers and institutions. 

Whilst the identification of suspicious authorship-for-sale researchers is one example, other opportunities are also available for institutions. If an institution is able to identify internal supervisor relationships, scientometric datasets can be used to identify collaborations that occur outside of the local research network, especially if there are no other researchers from the institution on the paper.  

More broadly, these analyses belong to a provenance-based approach to research integrity that interrogate where the research came from, rather than the text on the page. At the author level,  we might choose to supplement our author analysis by looking at the roles they play on publications via the use of the credit ontology \cite{10.1073/pnas.1715374115}, or for evidence that the authors have previously been awarded funding or have been listed on patents. 

Beyond authors there are also other networks of trust on a paper.  The presence of a grant identifier in the acknowledgements, for instance, signals that the research on which the publication is based has gone through peer review, and opens the authors up for sanctions by the funder should the work demonstrate misconduct \cite{10.1007/s11948-023-00459-9}. In a similar way, an ethics approval identifier, if legitimate, connects a paper back to an institutional review process. Provided that these links are audited by funders and institutions and cannot be made up (see, for example, \cite{10.1371/journal.pone.0278362}), these links are difficult to fabricate by paper mills, as they are markers of research that has evolved through time. 

Unlike technological approaches to the detection of fabricated content based on the body of a paper that can be implemented at the point of submission, provenance based approaches to the detection of paper mills must be implemented across the research community. In order for research provenance to be difficult to fake, provenance trails need to be systematically audited. A data-driven approach to research integrity is required, one involving collaboration between funders, institutions and publishers. In the example above, without enforcing that ethics approval identifiers and grant identifiers are recorded on papers in structured ways, institutions and funders cannot easily audit relevant publications. And in an era of generative AI, these provenance-based approaches are increasingly important.


\subsection{Calls to action}

There is some debate within the research integrity community on whether it is prudent to openly share approaches to paper-mill detection on the basis that, in doing so, paper-mill enterprises are able to adjust their strategies accordingly \cite{10.1038/d41586-022-04367-z}. While this position is understandable, we note that we have benefited significantly from the open Problematic Paper Screener dataset on tortured phrases, and the open investigations on PubPeer that it has seeded. Without access to this data, validation of our proposed technique would have been difficult if not impossible to carry out.  We believe that further transparency is important in the development of technologies that are strong enough to combat paper mills.

Finally, we believe that a window of opportunity exists to stop paper mills from creating false or ``engineered context'' for their fake papers by denying them the ability to place researchers with seemingly established researcher histories onto papers. The introduction of inexpensive access to large-language models and other emergent artificial intelligence tools is likely to make it all but impossible to detect a fraudulent manuscript solely using in-manuscript textual and graphical analyses.  We believe that this arms race will be expensive and difficult to win.  However, by making it difficult (or ideally impossible) for paper mills to create engineered context, we preserve a key detection mechanism which may become the critical tool in protecting this part of research integrity.

If we do not move quickly then the profiles in the authorship-for-sale network will become established---fake authors will look like real authors---and they will become increasingly difficult to detect with time. In addition, the need to overuse profiles will diminish as a greater number of new profiles within the authorship-for-sale network become established. This work will require increased activity from both institutions and publishers alike.

We believe that swift action is needed to change publishing methodologies to ensure that we make the establishment of these profiles more difficult. Requiring ORCiD is among the first steps.  It is critical to share information between publishers, institutions and funders so that the growth of the suspicious author network can be tracked and investigated. GDPR compliance makes direct data sharing challenging, however, constructing shared analysis based around shared scientometric datasets provides one path around this obstacle.  Holding information in silos will allow paper mills to establish themselves.  Winning alone is not winning at all---research integrity cannot and should not be viewed as a competitive advantage, but rather it is a common good.  Publishers, funders, and institutions that highlight their research integrity cases rather than hiding them or being ashamed of them, and that speak openly about their experiences should be praised rather than pilloried for their actions.  Only with this community-based transparency and spirit will we be able to protect the integrity of scholarly communication for future generations.




\section{Acknowledgements}
The authors would like to acknowledge the invaluable feedback we received during the drafting of this manuscript from Ann Campbell and Jeorg Sixt (Digital Science),  Dakota Murray (Northeastern University), and Guillaume Cabanac (University of Toulouse).

\section{Conflict of Interest}
Both authors are employed by Digital Science. \textit{Dimensions}, the database used in this analysis is owned and operated by Digital Science.

\section{Contributor's Roles}
\textbf{Simon J. Porter} [0000-0002-6151-8423]: Software, Conceptualization, Visualization, Methodology, Writing – Original Draft Preparation, Writing – Review \& Editing. \textbf{Leslie D. McIntosh} [0000-0002-3507-7468]: Conceptualization, Methodology, Writing – Original Draft Preparation, Writing – Review \& Editing. \textbf{Daniel W. Hook} [0000-0001-9746-1193]: Writing – Review \& Editing.

\section{Funding}
This research received no external funding, and was conducted as part of continuing thought leadership activities within Digital Science (https://ror.org/02ktfc112).

\section{Ethics Approval Statement}
No ethics approval was sought for this project.

\section{Data Availability Statement}
Whilst we have shared the queries and code that we have used to generate the analysis (doi: 10.6084/m9.figshare.24795864) we have not shared information on the individual researchers that we have identified. This choice does not reflect a desire to hold information secret, but rather a reflection on the damage to academic reputation that a false positive result could cause if presented out of context.

\bibliography{PaperMillPapers}

\begin{thebibliography}{48}
\providecommand{\natexlab}[1]{#1}
\providecommand{\url}[1]{\texttt{#1}}
\expandafter\ifx\csname urlstyle\endcsname\relax
  \providecommand{\doi}[1]{doi: #1}\else
  \providecommand{\doi}{doi: \begingroup \urlstyle{rm}\Url}\fi

\bibitem[10.(2015)]{10.1017/cbo9781316339831}
{Introduction to Random Graphs}.
\newblock 2015.
\newblock \doi{10.1017/cbo9781316339831}.

\bibitem[Abalkina(2023)]{10.1002/leap.1574}
A.~Abalkina.
\newblock {Publication and collaboration anomalies in academic papers originating from a paper mill: Evidence from a Russia‐based paper mill}.
\newblock \emph{Learned Publishing}, 36\penalty0 (4):\penalty0 689--702, 2023.
\newblock ISSN 0953-1513.
\newblock \doi{10.1002/leap.1574}.

\bibitem[Al-Marzouki et~al.(2005)Al-Marzouki, Evans, Marshall, and Roberts]{10.1136/bmj.331.7511.267}
S.~Al-Marzouki, S.~Evans, T.~Marshall, and I.~Roberts.
\newblock {Are these data real? Statistical methods for the detection of data fabrication in clinical trials}.
\newblock \emph{BMJ}, 331\penalty0 (7511):\penalty0 267, 2005.
\newblock ISSN 0959-8138.
\newblock \doi{10.1136/bmj.331.7511.267}.

\bibitem[Barbour and Stell(2020)]{10.7551/mitpress/11087.003.0015}
B.~Barbour and B.~M. Stell.
\newblock {Gaming the Metrics}.
\newblock pages 149--156, 2020.
\newblock \doi{10.7551/mitpress/11087.003.0015}.

\bibitem[Bik et~al.(2016)Bik, Casadevall, and Fang]{10.1128/mbio.00809-16}
E.~M. Bik, A.~Casadevall, and F.~C. Fang.
\newblock {The Prevalence of Inappropriate Image Duplication in Biomedical Research Publications}.
\newblock \emph{mBio}, 7\penalty0 (3):\penalty0 e00809--16, 2016.
\newblock ISSN 2161-2129.
\newblock \doi{10.1128/mbio.00809-16}.

\bibitem[Byrne and Christopher(2020)]{10.1002/1873-3468.13747}
J.~A. Byrne and J.~Christopher.
\newblock {Digital magic, or the dark arts of the 21st century—how can journals and peer reviewers detect manuscripts and publications from paper mills?}
\newblock \emph{FEBS Letters}, 594\penalty0 (4):\penalty0 583--589, 2020.
\newblock ISSN 0014-5793.
\newblock \doi{10.1002/1873-3468.13747}.

\bibitem[Cabanac et~al.(2022)Cabanac, Labbé, and Magazinov]{10.48550/arxiv.2210.04895}
G.~Cabanac, C.~Labbé, and A.~Magazinov.
\newblock {The 'Problematic Paper Screener' automatically selects suspect publications for post-publication (re)assessment}.
\newblock \emph{arXiv}, 2022.
\newblock \doi{10.48550/arxiv.2210.04895}.

\bibitem[Chawla(2023)]{vk}
D.~S. Chawla.
\newblock {Exposing a most unscrupulous journal}, 11 2023.
\newblock URL \url{https://cen.acs.org/policy/publishing/Exposing-unscrupulous-journal/101/i38}.

\bibitem[Combat-Bard(2023)]{10.31219/osf.io/wk3g7}
A.~Combat-Bard.
\newblock {How AI can help combat paper mills}.
\newblock 2023.
\newblock \doi{10.31219/osf.io/wk3g7}.

\bibitem[{Commission, Crime and Corruption}(2017)]{eg}
{Commission, Crime and Corruption}.
\newblock {Australia’s first criminal prosecution for research fraud}.
\newblock Technical report, 12 2017.
\newblock URL \url{https://services.anu.edu.au/files/guidance/Australias-first-criminal-prosecution-for-research-fraud-final.pdf}.

\bibitem[{Commission, Crime and Corruption}(2020)]{y9j}
{Commission, Crime and Corruption}.
\newblock {Reducing the risk of research fraud}.
\newblock Technical report, 7 2020.
\newblock URL \url{https://www.ccc.qld.gov.au/sites/default/files/Docs/Publications/CCC/Summary-audit-report-Reducing-the-risk-of-research-fraud-2020.pdf}.

\bibitem[Couzin-Frankel(2015)]{10.1126/science.349.6252.1036}
J.~Couzin-Frankel.
\newblock {PubPeer co-founder reveals identity—and new plans}.
\newblock \emph{Science}, 349\penalty0 (6252):\penalty0 1036--1036, 2015.
\newblock ISSN 0036-8075.
\newblock \doi{10.1126/science.349.6252.1036}.

\bibitem[Dondio et~al.(2019)Dondio, Casnici, Grimaldo, Gilbert, and Squazzoni]{10.1016/j.joi.2019.03.018}
P.~Dondio, N.~Casnici, F.~Grimaldo, N.~Gilbert, and F.~Squazzoni.
\newblock {The “invisible hand” of peer review: The implications of author-referee networks on peer review in a scholarly journal}.
\newblock \emph{Journal of Informetrics}, 13\penalty0 (2):\penalty0 708--716, 2019.
\newblock ISSN 1751-1577.
\newblock \doi{10.1016/j.joi.2019.03.018}.

\bibitem[Dunbar(1992)]{10.1016/0047-2484(92)90081-j}
R.~Dunbar.
\newblock {Neocortex size as a constraint on group size in primates}.
\newblock \emph{Journal of Human Evolution}, 22\penalty0 (6):\penalty0 469--493, 1992.
\newblock ISSN 0047-2484.
\newblock \doi{10.1016/0047-2484(92)90081-j}.

\bibitem[Editors(2022)]{10.1371/journal.pone.0278362}
T.~P.~O. Editors.
\newblock {Expression of Concern: Impact of smoking cessation, coffee and bread consumption on the intestinal microbial composition among Saudis: A cross-sectional study}.
\newblock \emph{PLOS ONE}, 17\penalty0 (12):\penalty0 e0278362, 2022.
\newblock \doi{10.1371/journal.pone.0278362}.

\bibitem[Edwards and Roy(2017)]{10.1089/ees.2016.0223}
M.~A. Edwards and S.~Roy.
\newblock {Academic Research in the 21st Century: Maintaining Scientific Integrity in a Climate of Perverse Incentives and Hypercompetition}.
\newblock \emph{Environmental Engineering Science}, 34\penalty0 (1):\penalty0 51--61, 2017.
\newblock ISSN 1092-8758.
\newblock \doi{10.1089/ees.2016.0223}.

\bibitem[Fanelli(2013)]{10.1371/journal.pmed.1001563}
D.~Fanelli.
\newblock {Why Growing Retractions Are (Mostly) a Good Sign}.
\newblock \emph{PLoS Medicine}, 10\penalty0 (12):\penalty0 e1001563, 2013.
\newblock ISSN 1549-1277.
\newblock \doi{10.1371/journal.pmed.1001563}.

\bibitem[Fanelli et~al.(2015)Fanelli, Costas, and Larivière]{10.1371/journal.pone.0127556}
D.~Fanelli, R.~Costas, and V.~Larivière.
\newblock {Misconduct Policies, Academic Culture and Career Stage, Not Gender or Pressures to Publish, Affect Scientific Integrity}.
\newblock \emph{PLoS ONE}, 10\penalty0 (6):\penalty0 e0127556, 2015.
\newblock \doi{10.1371/journal.pone.0127556}.

\bibitem[Fang et~al.(2012)Fang, Steen, and Casadevall]{10.1073/pnas.1212247109}
F.~C. Fang, R.~G. Steen, and A.~Casadevall.
\newblock {Misconduct accounts for the majority of retracted scientific publications}.
\newblock \emph{Proceedings of the National Academy of Sciences}, 109\penalty0 (42):\penalty0 17028--17033, 2012.
\newblock ISSN 0027-8424.
\newblock \doi{10.1073/pnas.1212247109}.

\bibitem[Flintoft et~al.(2023)Flintoft, MacCallum, Streeter, {Flanagan, Director of Research Integrity Strategy and Policy at Wiley Dave}, and Ferguson]{ao}
L.~Flintoft, S.~P. D. a. W. C.~J. MacCallum, D.~o. F. S. R. a. W.~M. Streeter, {Flanagan, Director of Research Integrity Strategy and Policy at Wiley Dave}, and S.~D. o. D. S. a. W.~L. Ferguson.
\newblock {Tackling publication manipulation at scale: Hindawi’s journey and lessons for academic publishing}.
\newblock Technical report, 12 2023.
\newblock URL \url{https://www.wiley.com/en-us/network/publishing/research-publishing/open-access/hindawi-publication-manipulation-whitepaper}.

\bibitem[FLYNN()]{40i}
J.~FLYNN.
\newblock {Guest Post — Addressing Paper Mills and a Way Forward for Journal Security}.
\newblock \emph{Scholarly Kitchen}.
\newblock URL \url{https://scholarlykitchen.sspnet.org/2023/04/04/guest-post-addressing-paper-mills-and-a-way-forward-for-journal-security/}.

\bibitem[Fortunato et~al.(2018)Fortunato, Bergstrom, Börner, Evans, Helbing, Milojević, Petersen, Radicchi, Sinatra, Uzzi, Vespignani, Waltman, Wang, and Barabási]{10.1126/science.aao0185}
S.~Fortunato, C.~T. Bergstrom, K.~Börner, J.~A. Evans, D.~Helbing, S.~Milojević, A.~M. Petersen, F.~Radicchi, R.~Sinatra, B.~Uzzi, A.~Vespignani, L.~Waltman, D.~Wang, and A.-L. Barabási.
\newblock {Science of science}.
\newblock \emph{Science}, 359\penalty0 (6379), 2018.
\newblock ISSN 0036-8075.
\newblock \doi{10.1126/science.aao0185}.

\bibitem[Graham(2022)]{10.1038/d41586-022-04367-z}
F.~Graham.
\newblock {Daily briefing: Publishers put paper-mill detectors to the test}.
\newblock \emph{Nature}, 2022.
\newblock ISSN 0028-0836.
\newblock \doi{10.1038/d41586-022-04367-z}.

\bibitem[Hagberg et~al.(2008)Hagberg, Schult, and Swart]{Hagberg.2008}
A.~A. Hagberg, D.~A. Schult, and P.~J. Swart.
\newblock {Exploring Network Structure, Dynamics, and Function using NetworkX}.
\newblock pages 11 -- 15, 2008.

\bibitem[Heusse and Cabanac(2022)]{10.1002/leap.1451}
M.~Heusse and G.~Cabanac.
\newblock {ORCID growth and field‐wise dynamics of adoption: A case study of the Toulouse scientific area}.
\newblock \emph{Learned Publishing}, 35\penalty0 (4):\penalty0 454--466, 2022.
\newblock ISSN 0953-1513.
\newblock \doi{10.1002/leap.1451}.

\bibitem[Hook and Porter(2021)]{10.3389/frma.2021.656233}
D.~W. Hook and S.~J. Porter.
\newblock {Scaling Scientometrics: Dimensions on Google BigQuery as an Infrastructure for Large-Scale Analysis}.
\newblock \emph{Frontiers in Research Metrics and Analytics}, 6:\penalty0 656233, 2021.
\newblock \doi{10.3389/frma.2021.656233}.

\bibitem[Hook et~al.(2018)Hook, Porter, and Herzog]{10.3389/frma.2018.00023}
D.~W. Hook, S.~J. Porter, and C.~Herzog.
\newblock {Dimensions: Building Context for Search and Evaluation}.
\newblock \emph{Frontiers in Research Metrics and Analytics}, 3:\penalty0 23, 2018.
\newblock \doi{10.3389/frma.2018.00023}.

\bibitem[Integrity(2018)]{q}
T.~C. f.~S. Integrity.
\newblock {The Retraction Watch Database}.
\newblock 1 2018.
\newblock ISSN 2692-465X.
\newblock URL \url{http://retractiondatabase.org/}.

\bibitem[Kim et~al.(2023)Kim, Kim, and Kim]{10.1177/01655515211018171}
J.~Kim, J.~Kim, and J.~Kim.
\newblock {Effect of Chinese characters on machine learning for Chinese author name disambiguation: A counterfactual evaluation}.
\newblock \emph{Journal of Information Science}, 49\penalty0 (3):\penalty0 711--725, 2023.
\newblock ISSN 0165-5515.
\newblock \doi{10.1177/01655515211018171}.

\bibitem[Kincaid(2022)]{ptw}
E.~Kincaid.
\newblock {How a tweet sparked an investigation that led to a PhD student leaving his program}, 8 2022.
\newblock URL \url{https://retractionwatch.com/2022/08/24/how-a-tweet-sparked-an-investigation-that-led-to-a-phd-student-leaving-his-program/}.

\bibitem[Lacetera and Zirulia(2011)]{10.1093/jleo/ewp031}
N.~Lacetera and L.~Zirulia.
\newblock {The Economics of Scientific Misconduct}.
\newblock \emph{The Journal of Law, Economics, \& Organization}, 27\penalty0 (3):\penalty0 568--603, 2011.
\newblock ISSN 8756-6222.
\newblock \doi{10.1093/jleo/ewp031}.

\bibitem[Liverpool(2023)]{10.1038/d41586-023-01780-w}
L.~Liverpool.
\newblock {AI intensifies fight against ‘paper mills’ that churn out fake research}.
\newblock \emph{Nature}, 618\penalty0 (7964):\penalty0 222--223, 2023.
\newblock ISSN 0028-0836.
\newblock \doi{10.1038/d41586-023-01780-w}.

\bibitem[Matveeva et~al.(2021)Matveeva, Sterligov, and Yudkevich]{10.1016/j.joi.2020.101110}
N.~Matveeva, I.~Sterligov, and M.~Yudkevich.
\newblock {The effect of Russian University Excellence Initiative on publications and collaboration patterns}.
\newblock \emph{Journal of Informetrics}, 15\penalty0 (1):\penalty0 101110, 2021.
\newblock ISSN 1751-1577.
\newblock \doi{10.1016/j.joi.2020.101110}.

\bibitem[McIntosh and Hudson(2024)]{10.6084/m9.figshare.24948789}
L.~D. McIntosh and C.~Hudson.
\newblock {Taxonomy of Retractions}.
\newblock Technical report, figshare, 1 2024.
\newblock URL \url{https://dx.doi.org/10.6084/m9.figshare.24948789}.

\bibitem[McNutt et~al.(2018)McNutt, Bradford, Drazen, Hanson, Howard, Jamieson, Kiermer, Marcus, Pope, Schekman, Swaminathan, Stang, and Verma]{10.1073/pnas.1715374115}
M.~K. McNutt, M.~Bradford, J.~M. Drazen, B.~Hanson, B.~Howard, K.~H. Jamieson, V.~Kiermer, E.~Marcus, B.~K. Pope, R.~Schekman, S.~Swaminathan, P.~J. Stang, and I.~M. Verma.
\newblock {Transparency in authors’ contributions and responsibilities to promote integrity in scientific publication}.
\newblock \emph{Proceedings of the National Academy of Sciences}, 115\penalty0 (11):\penalty0 2557--2560, 2018.
\newblock ISSN 0027-8424.
\newblock \doi{10.1073/pnas.1715374115}.

\bibitem[Montenegro(2023)]{Montenegro.2023}
A.~Montenegro.
\newblock {ORCID Public Data File 2023}.
\newblock 2023.
\newblock \doi{10.23640/07243.24204912.v1}.
\newblock URL \url{https://orcid.figshare.com/articles/dataset/ORCID\_Public\_Data\_File\_2023/24204912}.

\bibitem[Moylan and Kowalczuk(2016)]{10.1136/bmjopen-2016-012047}
E.~C. Moylan and M.~K. Kowalczuk.
\newblock {Why articles are retracted: a retrospective cross-sectional study of retraction notices at BioMed Central}.
\newblock \emph{BMJ Open}, 6\penalty0 (11):\penalty0 e012047, 2016.
\newblock ISSN 2044-6055.
\newblock \doi{10.1136/bmjopen-2016-012047}.

\bibitem[Noorden(2014)]{10.1038/nature.2014.16102}
R.~V. Noorden.
\newblock {The scientists who get credit for peer review}.
\newblock \emph{Nature}, 2014.
\newblock ISSN 0028-0836.
\newblock \doi{10.1038/nature.2014.16102}.

\bibitem[Nuijten and Wicherts(2023)]{10.31234/osf.io/bxau9}
M.~B. Nuijten and J.~M. Wicherts.
\newblock {The effectiveness of implementing statcheck in the peer review process to avoid statistical reporting errors}.
\newblock 2023.
\newblock \doi{10.31234/osf.io/bxau9}.

\bibitem[Oransky(2023)]{zqc}
I.~Oransky.
\newblock {The Retraction Watch Database becomes completely open – and RW becomes far more sustainable}, 9 2023.
\newblock URL \url{https://retractionwatch.com/2023/09/12/the-retraction-watch-database-becomes-completely-open-and-rw-becomes-far-more-sustainable/}.

\bibitem[O’Sullivan and Carr(2018)]{10.1177/1461444816686104}
P.~B. O’Sullivan and C.~T. Carr.
\newblock {Masspersonal communication: A model bridging the mass-interpersonal divide}.
\newblock \emph{New Media \& Society}, 20\penalty0 (3):\penalty0 1161--1180, 2018.
\newblock ISSN 1461-4448.
\newblock \doi{10.1177/1461444816686104}.

\bibitem[Park et~al.(2022)Park, West, Pathmendra, Favier, Stoeger, Capes-Davis, Cabanac, Labbé, and Byrne]{10.26508/lsa.202101203}
Y.~Park, R.~A. West, P.~Pathmendra, B.~Favier, T.~Stoeger, A.~Capes-Davis, G.~Cabanac, C.~Labbé, and J.~A. Byrne.
\newblock {Identification of human gene research articles with wrongly identified nucleotide sequences}.
\newblock \emph{Life Science Alliance}, 5\penalty0 (4):\penalty0 e202101203, 2022.
\newblock ISSN 2575-1077.
\newblock \doi{10.26508/lsa.202101203}.

\bibitem[Porter(2022)]{10.3389/frma.2022.779097}
S.~J. Porter.
\newblock {Measuring Research Information Citizenship Across ORCID Practice}.
\newblock \emph{Frontiers in Research Metrics and Analytics}, 7:\penalty0 779097, 2022.
\newblock \doi{10.3389/frma.2022.779097}.

\bibitem[Porter and Hook(2022)]{10.3389/frma.2022.835139}
S.~J. Porter and D.~W. Hook.
\newblock {Connecting Scientometrics: Dimensions as a Route to Broadening Context for Analyses}.
\newblock \emph{Frontiers in Research Metrics and Analytics}, 7:\penalty0 835139, 2022.
\newblock \doi{10.3389/frma.2022.835139}.

\bibitem[Silva(2021)]{10.1007/s11192-021-03996-x}
J.~A. T.~d. Silva.
\newblock {Abuse of ORCID’s weaknesses by authors who use paper mills}.
\newblock \emph{Scientometrics}, 126\penalty0 (7):\penalty0 6119--6125, 2021.
\newblock ISSN 0138-9130.
\newblock \doi{10.1007/s11192-021-03996-x}.

\bibitem[{STM, Committee on Publication Ethics and} and Services(2022)]{p6}
{STM, Committee on Publication Ethics and} and M.~P. Services.
\newblock {Paper mills research report and recommendations}.
\newblock Technical report, 6 2022.
\newblock URL \url{https://doi.org/10.24318/jtbG8IHL}.

\bibitem[Tang et~al.(2023)Tang, Wang, and Hu]{10.1007/s11948-023-00459-9}
L.~Tang, L.~Wang, and G.~Hu.
\newblock {Research Misconduct Investigations in China’s Science Funding System}.
\newblock \emph{Science and Engineering Ethics}, 29\penalty0 (6):\penalty0 39, 2023.
\newblock ISSN 1353-3452.
\newblock \doi{10.1007/s11948-023-00459-9}.

\bibitem[Watts and Strogatz(1998)]{10.1038/30918}
D.~J. Watts and S.~H. Strogatz.
\newblock {Collective dynamics of ‘small-world’ networks}.
\newblock \emph{Nature}, 393\penalty0 (6684):\penalty0 440--442, 1998.
\newblock ISSN 0028-0836.
\newblock \doi{10.1038/30918}.

\end{thebibliography}

\end{document}